\documentclass[superscriptaddress,twocolumn,prx, aps,preprintnumbers,notitlepage,longbibliography,floatfix]{revtex4-1}

\usepackage[latin9]{inputenc}
\usepackage{amsmath}
\usepackage{amssymb}
\usepackage{graphicx}
\usepackage{esint}
\usepackage[scr=boondoxo, scrscaled=1.05]{mathalfa}

\usepackage{hyperref}
\allowdisplaybreaks

\makeatletter

\pdfpageheight\paperheight
\pdfpagewidth\paperwidth

\pdfoutput=1

\usepackage{graphics}
\usepackage{mciteplus}
\mciteErrorOnUnknownfalse

\makeatother

\graphicspath{{./}{figures/}}
\usepackage{physics, multirow, array}
\usepackage[caption=false]{subfig}

\begin{document}

\preprint{FERMILAB-PUB-21-460-AE}

\title{Anomalies of Cosmic Anisotropy from Holographic Universality of Great-Circle Variance}

\author{Nathaniel Selub}
\affiliation{University of Chicago, 5640 South Ellis Ave., Chicago, IL 60637}

\author{Frederick Wehlen}
\affiliation{University of Chicago, 5640 South Ellis Ave., Chicago, IL 60637}

\author{Craig Hogan}
\affiliation{University of Chicago, 5640 South Ellis Ave., Chicago, IL 60637}
\affiliation{Fermi National Accelerator Laboratory, Batavia, IL 60510}

\author{Stephan S. Meyer}
\affiliation{University of Chicago, 5640 South Ellis Ave., Chicago, IL 60637}

\begin{abstract}

We examine all-sky cosmic microwave background (CMB) temperature maps on large angular scales to compare their consistency with two scenarios: the standard inflationary quantum picture, and a distribution constrained to have a universal variance of primordial curvature perturbations on great circles. The latter symmetry is not a property of standard quantum inflation, but may be a symmetry of holographic models with causal quantum coherence on null surfaces. Since the variation of great-circle variance is dominated by the largest angular scale modes, in the latter case the amplitude and direction of the unobserved intrinsic dipole (that is, the $\ell=1$ harmonics) can be estimated from measured $\ell = 2, 3$ harmonics by minimizing the variance of great-circle variances including only $\ell =1, 2, 3$ modes. It is found that including the estimated intrinsic dipole leads to a nearly-null angular correlation function over a wide range of angles, in agreement with a null anti-hemispherical symmetry independently motivated by holographic causal arguments, but highly anomalous in standard cosmology. Simulations are used here to show that simultaneously imposing the constraints of universal great-circle variance and the vanishing of the angular correlation function over a wide range of angles tends to require patterns that are unusual in the standard picture, such as anomalously high sectorality of the $\ell = 3$ components, and a close alignment of principal axes of $\ell=2$ and $\ell = 3$ components, that have been previously noted on the actual sky. The precision of these results appears to be primarily limited by errors introduced by models of Galactic foregrounds.

\end{abstract}

\maketitle

\section{Introduction}

According to the standard cosmological model, the pattern of temperature anisotropy in the cosmic microwave background (CMB) on large angular scales approximates that of an invariant primordial scalar potential perturbation, $\Delta$, on the surface of last scattering. Correlations of the largest angular structures provide our most direct measurement of quantum processes that lay down primordial perturbations at the earliest times.

Although the standard cosmological model agrees extraordinarily well with the angular power spectrum of the observed cosmic microwave background at scales smaller than a few degrees \citep{Akrami:2018vks}, the angular distribution on large angular scales departs considerably from expectations \citep{de_oliveira-costa,WMAPanomalies,2016A&A...594A..16P,2016CQGra..33r4001S, Planck-2018-Isotropy}. Among the well-studied ``anomalies'' are a remarkably small value of the two-point angular correlation function at large angles \citep{1992ApJ...396L..13W,1994ApJ...436..423B,Copi2010}, including a near-vanishing value at $\Theta = 90^\circ$ \citep{Hagimoto_2020}; a highly planar octupole ($\ell=3$) component closely aligned with the quadrupole ($\ell=2$) \citep{2015MNRAS.449.3458C}; and a tendency for odd multipole moments to have more fluctuation power than even multipole moments \citep{PhysRevD.82.063002}. In the standard model of cosmology, where the $a_{\ell m}$ components of multipole moments are predicted to be independent Gaussian random variables, the vanishing of the two-point angular correlation function near $\Theta = 90^\circ$ is anomalous at a sub-$0.1\%$ level, while the other features are each anomalous at about a one-percent level.
In the standard view, these features are all statistical flukes: the chance of all of these anomalies occurring simultaneously is extremely low, and there is no particular relationship between them.

An alternative model of inflationary quantum states, which invokes holographic coherence on causal inflationary horizons, creates an opportunity to connect these anomalies via new, fundamental physical symmetries \citep{PhysRevD.99.063531,Hogan_2020,hogan2021angular}. Although standard cosmology is not significantly modified at angular scales less than a few degrees, large angles are affected by radically new correlations imposed by the emergence of space-time locality and causality from a quantum system. Viewed in this light, some anomalies can be seen as hints of deeper underlying symmetries from quantum gravity that have not been included in the standard scenario.

Predicted correlations of primordial perturbations depend on untested assumptions about quantum locality and coherence of geometrical states. In the standard quantum inflation model, zero-point fluctuations of coherently quantized scalar field modes produce a universal 3D power spectrum of primordial scalar curvature plane-wave perturbations on surfaces of constant time. By contrast, if quantum space-time is holographic in the sense suggested by some formal studies \citep{Banks:2018ypk,Banks:2020dus,Banks:2021jwj}, non-localized quantum states of geometry may be coherent on spherical bounding surfaces of causal diamonds. In holographic inflation, these spherical surfaces are defined by causally-coherent inflationary horizons around every observer, which can lead to a 2D power spectrum of relic perturbations on comoving spheres that obeys symmetries in the angular domain with no cosmic variance. A model of this kind creates new nonlocal (but causal) correlations on large angular scales that seem conspiratorial in the standard picture. They arise from quantum non-locality on spherical causal horizon surfaces, an effect not included in a standard scenario based on coherent plane-symmetric quantum field states.

The holographic scenario naturally leads to the conjecture that the two-point angular correlation function of $\Delta$ on 2-spheres approximates a function constrained by causal entanglements of horizons. This hypothesis appears to agree with symmetries of the observed angular correlation on large angular scales, in particular, the near-vanishing of angular correlation at angular separation $\Theta\geqslant 90^\circ$ \citep{hogan2021angular}. 

However, symmetries of the two-point angular correlation function do not directly address anomalies of shape and orientation determined by the phases of spherical harmonic components, which do not affect the two-point angular correlation function. In principle, holographic quantum gravity can also create surprising symmetries that constrain relationships between these components, again from causal coherence of quantum-gravitational states on null surfaces. For example, according to one form of the holographic scenario \cite{hogan2021angular}, local invariant scalar curvature at a point is determined by a coherent quantum state of its past light cone, while relationships with a definite spatial direction are governed by the state of a normal null plane. In this scenario, great circles have a special causal significance, since they delineate boundaries of coherent causal entanglement. A great circle on a sphere is the intersection of two null surfaces that carry relational information, a null plane (or infinitely large null sphere), and a light cone with an apex at the center of the sphere. These simple geometrical relationships suggest that the emergence of space-time locality may lead to universal properties of relic curvature variance on great circles. 
 In this case, it could be that some well known shape and alignment anomalies of the cosmic pattern, which depend on phases of multipoles that do not affect the two-point correlation function, are consequences of a new, fundamental physical symmetry.
Here, we investigate consequences of the hypothesis that {\it any great circle has the same one-point correlation function as any sphere}, or equivalently, {\it the one-point variance $\langle\Delta^2\rangle$ has a universal value for any great circle}. 

This hypothesis cannot be tested directly, even in principle: it depends on fine scale variations that cannot be measured. On the other hand, the variance among great circles in a given realization is dominated by low-order, large-angle modes, so
we can use a matched-filter approach to compare hypotheses. 
That is, the total variance increases with resolution or harmonic cutoff scale $\ell$, so it is dominated by the smallest angular scales, but its variation is dominated by large angular scales, so the effect of a symmetry manifests most conspicuously in skies filtered at low $\ell$. Here, we ask the question: for only $\ell\le 3$, do the great circles on the real sky more closely resemble the standard prediction, or a model with constant-variance symmetry of great circles?
Our approach is not optimized, but is a simple way to reduce the noise from small scales.

%Holographic symmetries of CMB temperature must be tested indirectly because exact angular symmetries of $\Delta$ are contaminated by measurement limitations.

%An important measurement limitation at large angles is Galactic foreground emission, which is imperfectly subtracted and leaves large-scale, non-primordial patterns in the maps. This aspect of the measurement can be improved with better data and models of all sky anisotropy over a wide range of frequencies.

One of the components in this analysis is the intrinsic cosmic dipole ($\ell=1$) anisotropy, which is not measurable, even in principle, from CMB data alone. Since the measured cosmic dipole is dominated by a kinetic dipole from our peculiar velocity that is orders of magnitude larger than the predicted intrinsic dipole, it is not possible to directly measure the orientation and amplitude of the true intrinsic dipole \citep{Ferreira_2021,Nadolny_2021}. Yet, the intrinsic dipole affects all-sky primordial symmetries: for example, it changes the two-point angular correlation function \cite{hogan2021angular} by adding a term proportional to $\cos(\Theta),$ with a positive coefficient, and also contributes to variances on great circles.

In the standard picture, there is no correlation among different spherical harmonics; the amplitudes of individual harmonic components are independent random variables. By contrast, holographic causal entanglement among spherical horizons requires causal correlations among low order spherical harmonic components. A causal constraint on the variation of great circle variances implies a relationship between the intrinsic (unmeasured) $\ell=1$ dipole component and the measured $\ell=2$ and $\ell=3$ components. Since the dipole is characterized by just three numbers, an amplitude and a direction, it is possible to estimate them from the measured components by minimizing the variation of filtered great-circle variances.

Thus, in principle the anomalous effects of a universal great-circle symmetry can be detected with measurements on a single sphere, such as the large-scale surface mapped by the CMB. In this paper, we use simulated skies to explore whether there is evidence for the posited symmetry in the real sky. We study its consistency with the data, with predicted realizations of the standard model, with causal symmetries of the two-point angular correlation function, and with measured anomalies of multipole shape and orientation.

\begin{figure*}[hbt]
\subfloat{\includegraphics[width=0.9\textwidth]{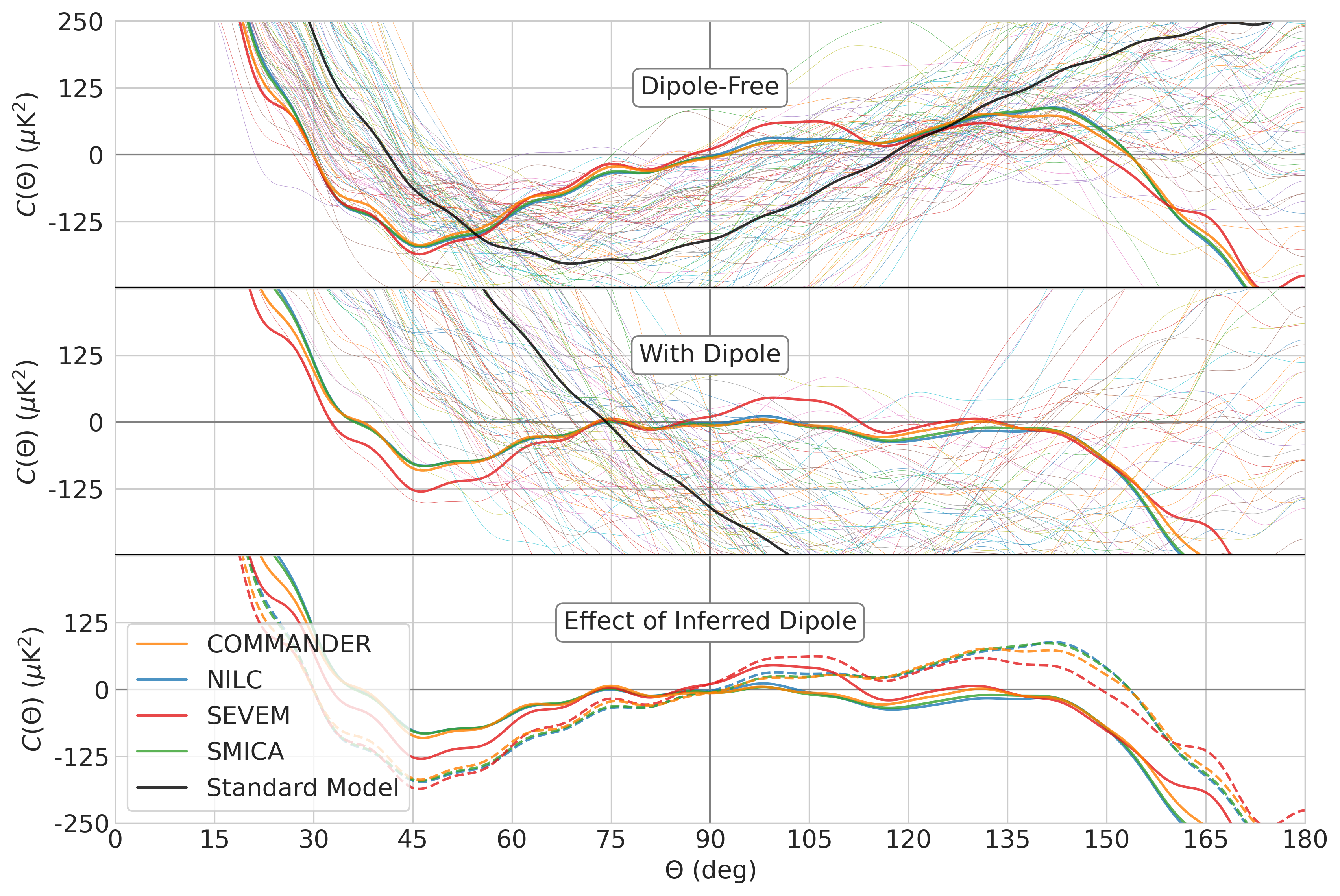}}
%\plotone{corrfunc.png}
\caption{Measurements and predictions for the two-point angular correlation function $C(\Theta)$, evaluated with a sharp cutoff $\ell_{max} = 30$. \emph{[Top]}: The two-point angular correlation functions of the dipole-free \textsl{Planck} maps (thick colored lines) are compared to those of the dipole-free theoretical standard model expectation (thick black line) and its realizations (thin lines). \emph{[Middle]}: The two-point angular correlation functions of the \textsl{Planck} maps with their inferred dipoles (thick colored lines) are compared to those of the theoretical standard model expectation with its expected dipole (thick black line), and its realizations with dipoles included (thin lines). The contrast between the two-point angular correlation functions of the \textsl{Planck} maps with their inferred dipoles and the two-point angular correlation functions predicted by the standard model illustrates the atypical nature of the inferred real sky in the standard model. \emph{[Bottom]}: $C(\Theta)$ for each \textsl{Planck} map with its inferred dipole (solid lines) is compared against $C(\Theta)$ for each dipole-free \textsl{Planck} map (dashed lines). This shows how the addition of the dipole inferred from the posited great-circle variance symmetry causes $C(\Theta)$ to approximately vanish on the angular interval $\Theta = [90^\circ, 135^\circ]$. \label{fig:corrfunc}}
\end{figure*}

\begin{figure*}[hbt]
\subfloat{\includegraphics[width=0.9\textwidth]{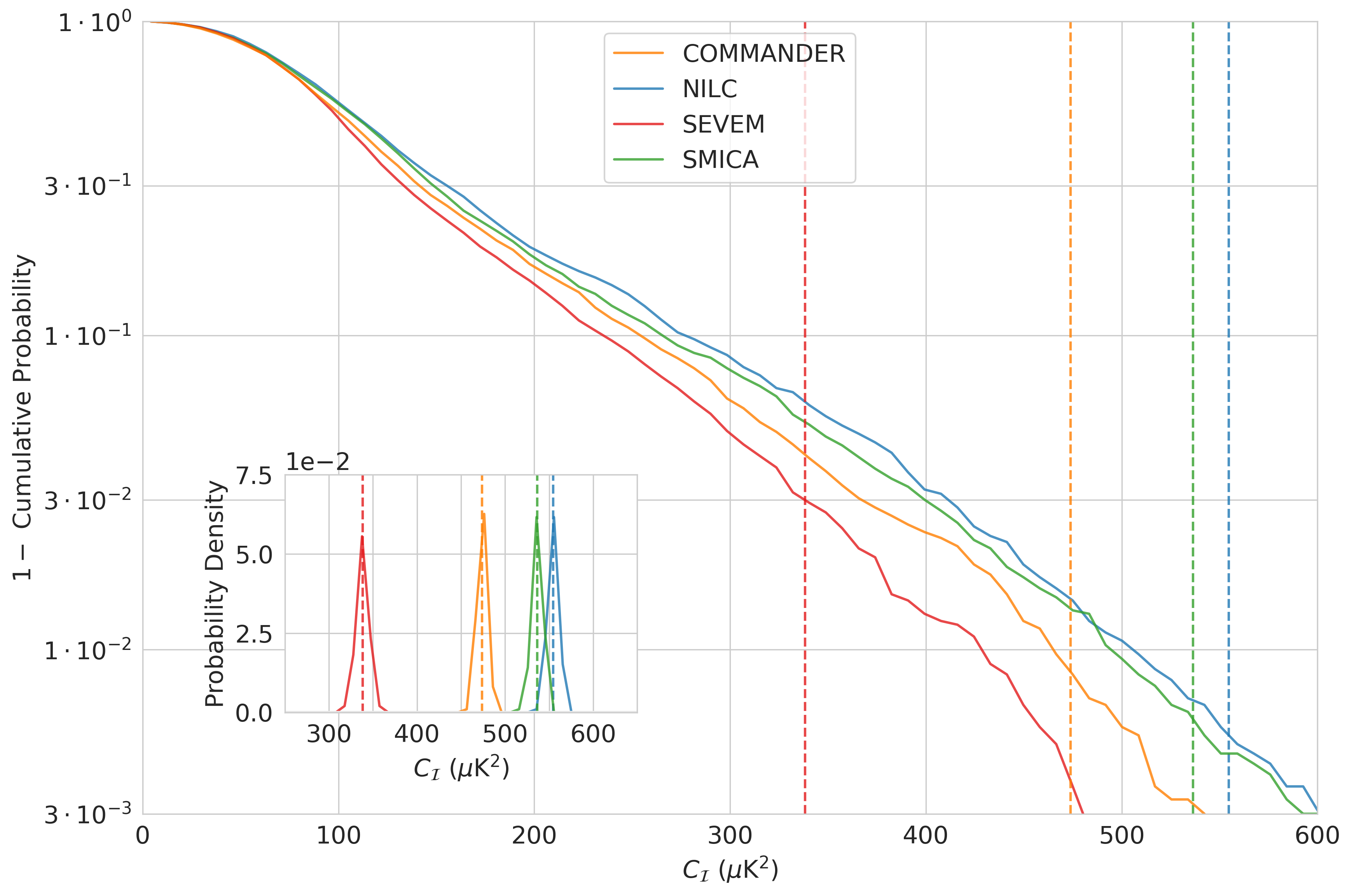}}
%\plotone{amplitude.png}
\caption{The inferred dipole amplitude $C_\mathcal{I}$ for each $\ell \leqslant 3$ \textsl{Planck} map (vertical lines) is compared with $C_\mathcal{I}$ for $3{,}000$ same-spectrum realizations of that map. Each $\ell \leqslant 3$ \textsl{Planck} map's set of realizations is used to create a reverse cumulative distribution function, and the cumulative probability density is the proportion of same-spectrum realizations of each $\ell \leqslant 3$ \textsl{Planck} map with values exceeding $C_\mathcal{I}$. The inset shows the spread in values of $C_\mathcal{I}$ for $100$ different $\mathcal{D}_\mathcal{I}$ for each $\ell \leqslant 3$ \textsl{Planck} map, calculated from $100$ different sets of $20{,}000$ uniformly distributed random great circles. In both plots, the vertical line indicates the mean value of $C_\mathcal{I}$ from each $\ell \leqslant 3$ \textsl{Planck} map's spread of inferred dipole amplitudes. Thus, the inferred intrinsic dipole that best fulfills the candidate symmetry of constant variance along great circles approximately cancels the two-point angular correlation function over the interval $\Theta = [90^\circ, 135^\circ]$ in only a very small subset of standard model realizations. \label{fig:amplitude}}
\end{figure*}

\subsection{Overview of results}

Remarkably, we find that the amplitude of the intrinsic dipole that optimizes constant variance on great circles for measured sky maps agrees with the value \cite{hogan2021angular} that approximately cancels the measured two-point angular correlation function over a wide angular interval. With this dipole, the ``true'' angular correlation in the range $\Theta = [\gtrsim 90^\circ, \lesssim 135^\circ]$ appears to be consistent with zero up to foreground subtraction errors, as expected from anti-hemispherical causal relationships in a holographic model. We show with simulations that this result is highly anomalous in standard cosmology (Fig. \ref{fig:corrfunc}), where it requires a fine-tuning of spectral amplitudes over a wide range of $\ell$ in the harmonic domain. Furthermore, we show that agreement between the intrinsic dipole that optimizes constant variance on great circles and the intrinsic dipole that approximately cancels the measured two-point angular correlation function on large angular scales is very rare for simulated combinations of $\ell = 2, 3$ that do not have the same anomalous features as those of measured sky maps (Fig. \ref{fig:amplitude}).

According to our holographic hypothesis, higher harmonics ought to smooth out variations of great-circle variance from lower harmonics. This behavior is manifest in the lowest harmonics of the real sky. Figs. (\ref{fig:mollweide}) and (\ref{fig:equirectangular}) show measured $\ell=2$ and $\ell=3$ distributions, along with an added ``inferred'' intrinsic dipole that comes closest to constant variance on great circles. The Mollweide projections (Fig. \ref{fig:mollweide}) use standard Galactic coordinates, while the axis adopted for the equirectangular (Cartesian) projections (Fig. \ref{fig:equirectangular}) is that defined by the inferred dipole. These projections show the high sectorality (see Eq. \eqref{eq:sect}) and alignment of the $\ell=2$ and $\ell=3$ distributions, and help visualize how these properties reduce variations of great-circle variance.

Thus, simultaneously imposing a small variance of great-circle variances on the $\ell \leqslant 3$ sky and the vanishing of angular correlation in the range $\Theta = [\gtrsim 90^\circ, \lesssim 135^\circ]$, for the fixed angular power spectrum as measured, tends to require previously noted anomalies of multipole shape. In particular, the principal axis of the intrinsic dipole tends to be closely aligned with those of the quadrupole and octupole components, and the octupole pattern tends to be highly sectoral. 

% Simulations show that the measured distribution of great circle variances, as well as the inferred dipole amplitude and total large-angle correlation, are atypical compared with random realizations predicted by the standard quantum model of inflation
% (Fig. \ref{fig:standard}). 

\section{Framework and Definitions}

\subsection{Variances of great circles}

The standard cosmological model predicts that on large angular scales, in the Sachs-Wolfe approximation, the $a_{\ell m}$ coefficients of the spherical-harmonic decomposition of the CMB sky are Gaussian distributed random variables with fixed variances. In this view, the observed power spectrum is one of many realizations of the model, which is statistical in nature.

A holographic model of inflation may produce an angular power spectrum with much less variation.
In the limiting case of a universal spectrum, the total variance of fluctuation power, proportional to the summed variances of $a_{\ell m}$'s at each $\ell$ in our sky, is the only one that could result from any physical realization. In this case, the phases of the $a_{\ell m}$ coefficients carry all the variation between different realizations, and contain additional information that encode higher-order correlations.

Since we have only one realization of the sky, we seek evidence for particular higher-order symmetries by comparing anomalies of our sky to those of simulated realizations that have the same angular power spectrum. We posit a simple symmetry, that \emph{the variance of $\Delta$ on great circles is a constant}, or equivalently, that \emph{ the variance of great-circle variances $\sigma^2_V$ vanishes}, i.e.,
\begin{equation}
\sigma^2_V = 0.
\end{equation}

Total variance is dominated by the smallest angular scale structure, so there is relatively little difference between holographic and standard pictures at high angular resolution. The clearest signature of a new symmetry in $\sigma^2_V$ is that the sky should have anomalously low $\sigma^2_V$ at low resolution, that is, including multipole moments of the spherical-harmonic decomposition of $\delta T$ only up to some small $\ell$ cutoff. Here, we study the great-circle variances from only the lowest multipole components, $\ell = 1, 2, 3$.
 
On angular scales larger than a few degrees $\ell \lesssim 30$, the cosmological part of the anisotropy is dominated by the Sachs-Wolfe effect. On the largest scales, the structure approximates a direct map of the primordial scalar curvature $\Delta$, so it approximately preserves the symmetries of the original pattern created by quantum inflation. Here, we include higher resolution structure up to $\ell = 30$ when evaluating the effect of the $\ell = 1, 2, 3$ components on the overall correlation function. 

\subsection{Harmonic analysis}

Given a CMB temperature map $\mathcal{M}$, let $\delta T (\hat{\Omega})$ denote the CMB temperature deviation from the overall mean as a function of position on the sky, where $\hat{\Omega}$ is determined by the polar angle $\theta$ and azimuthal angle $\varphi$,
\begin{equation}
\hat{\Omega} = (\theta, \varphi).
\end{equation}
$\delta T$ can be expanded as a linear combination of spherical harmonics $Y_{\ell m}$,
\begin{equation}
\delta T(\theta, \varphi) = \sum_{\ell = 0}^\infty \sum_{m=-\ell}^{m=+\ell}a_{\ell m} Y_{\ell m}(\theta, \varphi),
\end{equation}
where the $a_{\ell m}$ are harmonic coefficients constrained by the relationship
\begin{equation} \label{eq:constr}
a_{\ell,-m} = (-1)^m(a_{\ell, m})^*
\end{equation}
since $\delta T$ is real. The angular power spectrum $C_\ell$ of $\delta T$ is given by
\begin{equation} \label{eq:cl}
C_\ell = \frac{1}{2 \ell + 1} \sum_{m=-\ell}^{m=+\ell} \left|a_{\ell m}\right|^2,
\end{equation}
which we can use to analytically compute the two-point angular correlation function $C(\Theta)$ of $\delta T$ as
\begin{equation}
C(\Theta) = \frac{1}{4 \pi} \sum_{\ell = 0}^\infty (2\ell + 1) C_\ell P_\ell(\cos \Theta),\label{eq:corr}
\end{equation}
where the $P_\ell$ are the Legendre polynomials.

We define the dipole component $\mathcal{D}(\theta, \varphi)$ of $\delta T$ as
\begin{equation}
\mathcal{D}(\theta, \varphi) = \sum_{m=-1}^{m=+1}a_{1, m} Y_{1, m}(\theta, \varphi).
\end{equation}
$\mathcal{D}$ is uniquely determined by the location $\hat{\Omega}$ of its positive pole and the amplitude $C_1$ of its angular power spectrum term. Additionally, given the map $\mathcal{M}$, we define the map $\mathcal{M}'$ as a version of $\mathcal{M}$ with its dipole set to zero, and we define the $\ell \leqslant n$ version $\mathcal{M}_n$ of $\mathcal{M}$ as a version of $\mathcal{M}$ whose $\delta T$ has $C_\ell = 0$ for $\ell > n$.

Let $\hat{n}$ denote the direction of the $z$-axis of the coordinate system in which the spherical-harmonic decomposition of $\delta T$ was performed. Then, we define the \emph{sectorality} $\mathcal{S}_\ell(\hat{n})$ of each multipole in this coordinate system as
\begin{equation} \label{eq:sect}
\mathcal{S}_\ell(\hat{n}) = \sum_{m = -\ell, \ell} m^2 \abs{a_{\ell m}}^2.
\end{equation}
For each multipole, except $\ell = 0$, we define the \emph{axis of maximum sectorality} $\hat{N}_\ell$ as the direction of the $z$-axis in the coordinate system in which $\mathcal{S}_\ell$ is maximized for that particular multipole. Sectorality is similar to another coordinate system dependent statistic, planarity, which is defined as
\begin{equation}
\sum_{m = - \ell}^{m = + \ell} m^2 \abs{a_{\ell m}}^2.
\end{equation}
For low $\ell$ multipoles that are highly sectoral, the axis that maximizes planarity is very close to the axis of maximum sectorality, since the maximization of planarity is dominated by the $m = \pm \ell$ terms of the sum in the statistic's definition.

\subsection{Variance of great-circle variances}

For each great circle $c_i$, let $\delta T_i$ denote the distribution of $\delta T(\hat{\Omega})$ on $c_i$. We define the \emph{great-circle variance} $V_i$ for each $c_i$ as
\begin{equation}
V_i \equiv \sigma^2(\delta T_i) = \frac{1}{2 \pi} \oint_{c_i} \left[\delta T(\hat{\Omega}) - \mu(\delta T_i)\right]^2 d\hat{\Omega},
\end{equation}
where 
\begin{equation}
\mu(\delta T_i) = \frac{1}{2\pi} \oint_{c_i} \delta T(\hat{\Omega}') \,d\hat{\Omega}'.
\end{equation}
There exists a distribution $V$ of all such $V_i$ calculated for each $c_i$, and we define the \emph{variance of great-circle variances} $\sigma^2_V$ as
\begin{equation}
\sigma^2_V \equiv \sigma^2(V).
\end{equation}

\subsection{Inferred dipole}

We define the \emph{inferred dipole of $\mathcal{M}$} as the dipole $\mathcal{D}_\mathcal{I}(\mathcal{M})$ that minimizes $\sigma^2_V(\mathcal{M}_3' + \mathcal{D})$, i.e., $\mathcal{D}_\mathcal{I}(\mathcal{M})$ satisfies
\begin{equation}
\sigma^2_V(\mathcal{M}_3' + \mathcal{D}_\mathcal{I}(\mathcal{M})) \leqslant \sigma^2_V(\mathcal{M}_3' + \mathcal{D})
\end{equation}
for all possible $\mathcal{D}$. The location of the positive pole and the amplitude of the angular power spectrum term of $\mathcal{D}_\mathcal{I}(\mathcal{M})$ are denoted as $\hat{\Omega}_\mathcal{I}(\mathcal{M})$ and $C_\mathcal{I}(\mathcal{M})$, respectively. Next, we define the \emph{inferred version} of $\mathcal{M}$ (inferred $\mathcal{M}$) as the map
\begin{equation}
\mathcal{I}(\mathcal{M}) \equiv \mathcal{M}' + \mathcal{D}_\mathcal{I}(\mathcal{M}).
\end{equation}
Finally, we define a \emph{natural coordinate system} of $\mathcal{M}$ as a coordinate system in which the positive pole of $\mathcal{D}_\mathcal{I}$ coincides with $\hat{z}$ and the $m=\ell$ harmonic coefficient $a_{3,3}$ of its octupole moment is real and positive. Because $Y_{3 3} \propto e^{3i\phi}$, there are three such coordinate systems, related by rotations of $2\pi/3$ about the positive pole of $\mathcal{D}_\mathcal{I}$. Since this choice of orientation is immaterial, we choose one of these three possible coordinate systems and refer to it as \emph{the} natural coordinate system for simplicity. Fig. \ref{fig:equirectangular} indicates the chosen natural coordinate system.
\subsection{Standard Model realizations}

Let $C_\ell^\text{\,SM}$ denote the theoretical power spectrum predicted by the standard model. We define a \emph{standard model realization} as a CMB temperature map whose harmonic coefficients $\tilde{a}_{\ell m}$ are generated in the following manner. For each $\ell$, one number is chosen from a normal distribution with zero mean and variance $C_\ell^\text{\,SM}$, and $2\ell$ numbers are chosen from a normal distribution with zero mean and variance $C_\ell^\text{\,SM} / 2$. Then, these numbers are respectively assigned to one real coefficient $\tilde{a}_{\ell, 0}$ and $\ell$ complex coefficients $\tilde{a}_{\ell m}$ for $m = 1, 2, \dots, \ell$, composed of $\ell$ real and $\ell$ imaginary components. By Eq. \eqref{eq:constr}, the coefficients $\tilde{a}_{\ell m}$ for $m=-1,-2,\dots,-\ell$ are also determined.

\subsection{Standard Model dipole}

A standard model intrinsic cosmic dipole $\mathcal{D}_\mathcal{S}$ is generated by choosing one number from a normal distribution with zero mean and variance $C_1^\text{\,SM}$ and two numbers from a normal distribution with zero mean and variance $C_1^\text{\,SM} / 2$. These numbers are respectively assigned to the real coefficient $\beta_{1,0}$ and the real and imaginary components of the complex coefficient $\beta_{1,1}$. By Eq. \eqref{eq:constr}, the coefficient $\beta_{1,-1}$ is also determined.

\begin{figure*}[hbt]
\centering
\begin{tabular}{cc}
\multirow{-7.6}[0]{*}{\subfloat{\includegraphics[width=0.32\textwidth]{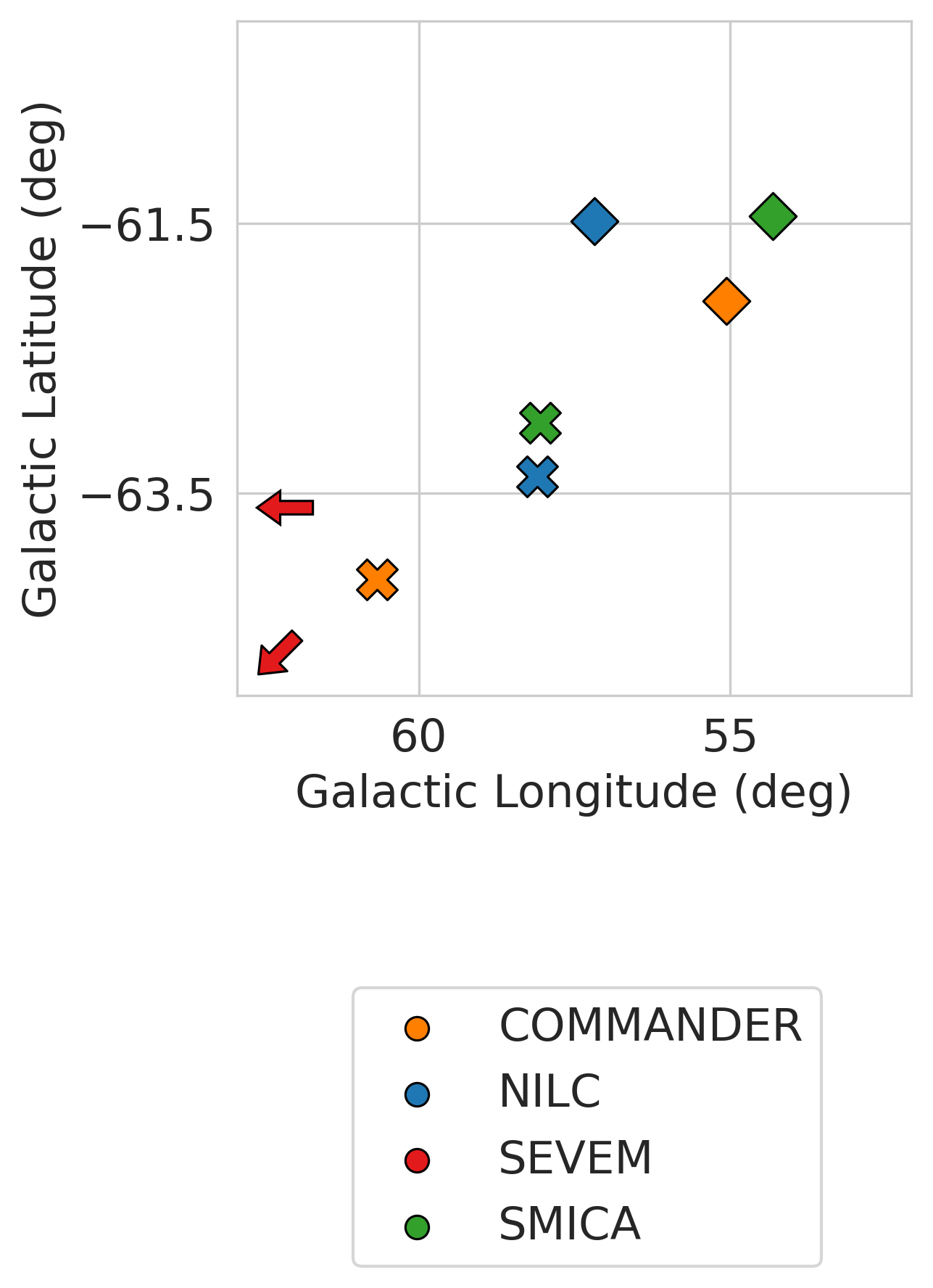}}} &
\subfloat{\includegraphics[width=0.6\textwidth]{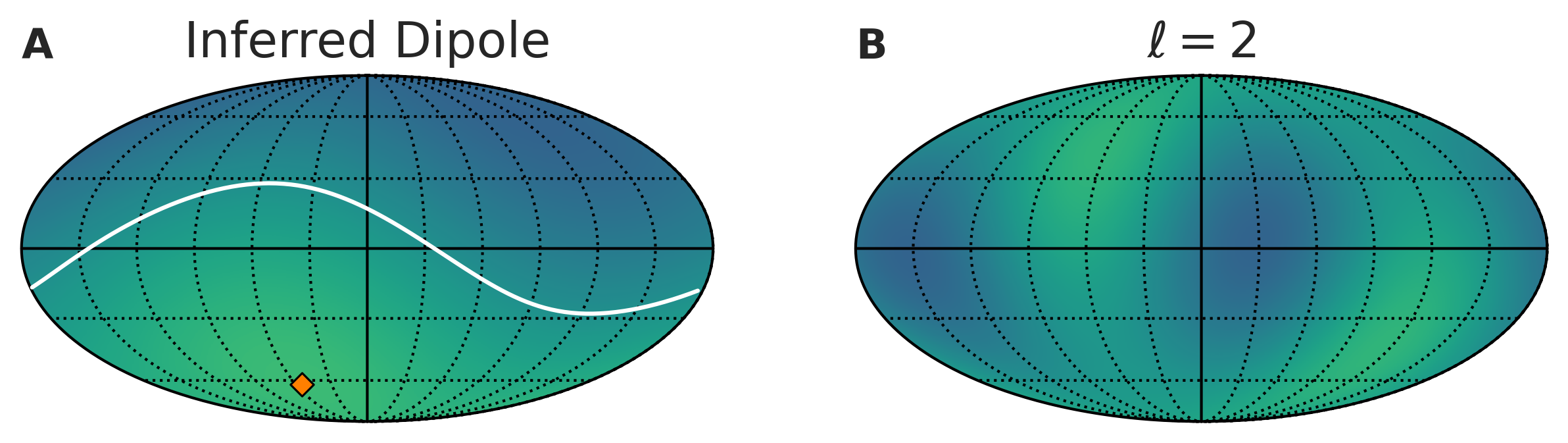}}\\
\subfloat{}&
\subfloat{\includegraphics[width=0.6\textwidth]{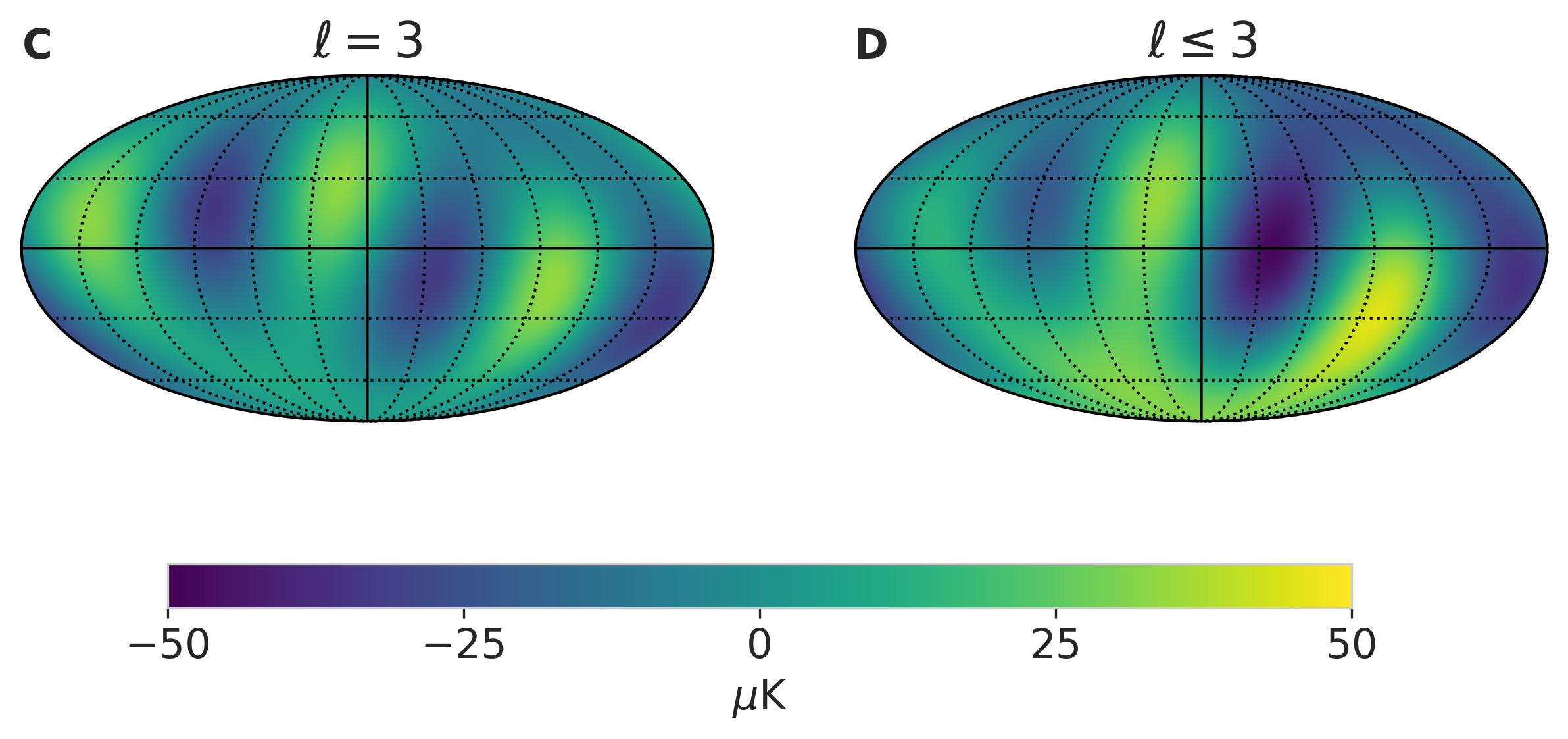}}
\end{tabular}
\protect\caption{On the left, an equirectangular (Cartesian) projection of a region of the sky showing the inferred dipole axis $\hat{\Omega}_\mathcal{I}$ (diamonds) and octupole axis of maximum sectorality $\hat{N}_3$ (crosses) in Galactic coordinates for the \textsl{Planck} pipelines Commander, NILC, and SMICA. The axes for SEVEM lie outside the region and are instead pointed to with arrows; its $\hat{\Omega}_\mathcal{I}$ is located at $(l, b) \approx (83.1^\circ, -63.6^\circ)$, and its $\hat{N}_3$ is located at $(l, b) \approx (67.1^\circ, -66.4^\circ)$, where $l$ and $b$ denote Galactic longitude and latitude, respectively. On the right (labeled A to D), four Mollweide projections of the low $\ell$ harmonic components of the inferred version of Commander are shown in Galactic coordinates. \emph{[Panel A]}: The inferred dipole $(\ell = 1)$ for Commander, annotated with the inferred dipole axis $\hat{\Omega}_\mathcal{I}$ (orange diamond) and inferred dipole equator (white line), which are used to define the natural coordinate system of Commander. \emph{[Panel B]}: The quadrupole $(\ell = 2)$ of Commander. \emph{[Panel C]}: The octupole $(\ell = 3)$ of Commander. \emph{[Panel D]}: The inferred version of $\ell \leqslant 3$ Commander, obtained by summing its octupole, quadrupole, and inferred dipole. \label{fig:mollweide}}
\end{figure*}

\begin{figure*}
\begin{centering}
\subfloat{\includegraphics[width=0.9\textwidth]{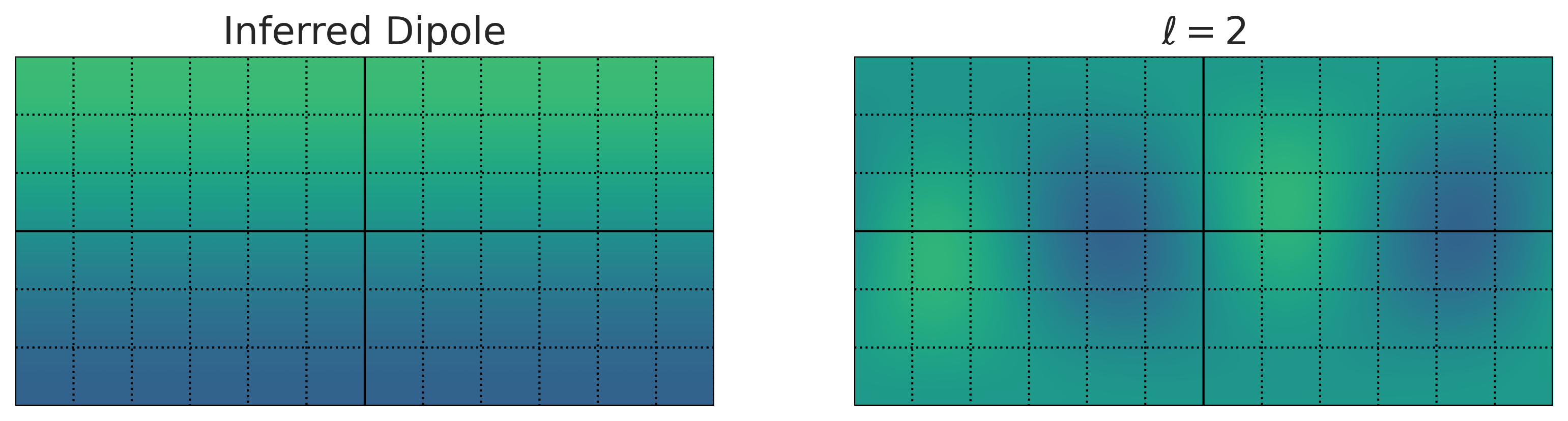}} \\
\subfloat{\includegraphics[width=0.9\textwidth]{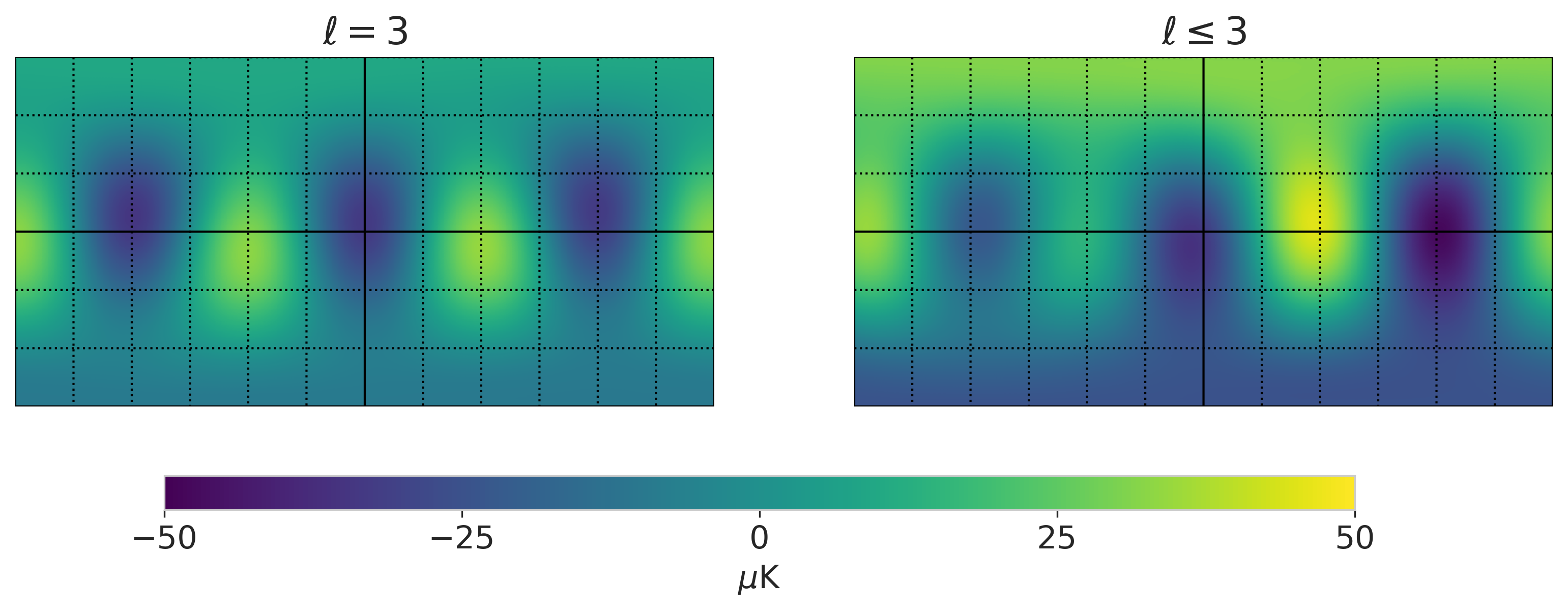}}
\par\end{centering}
\protect\caption{Equirectangular (Cartesian) projections of the low $\ell$ harmonic components of the inferred version of Commander, shown in its natural coordinate system. In the natural coordinate system, the positive pole of $\mathcal{D}_\mathcal{I}$ coincides with $\hat{z}$ and the $m=\ell$ harmonic coefficient $a_{3,3}$ of the octupole moment is real and positive. \emph{[Top Left]}: The inferred dipole for Commander. \emph{[Top Right]}: The quadrupole $(\ell = 2)$ of Commander. \emph{[Bottom Left]}: The octupole $(\ell = 3)$ of Commander. \emph{[Bottom Right]}: The inferred version of $\ell \leqslant 3$ Commander, obtained by summing its octupole, quadrupole, and inferred dipole. These maps show how the alignment and sectorality of the $(\ell = 2)$ and $(\ell = 3)$ components are related to the inferred intrinsic dipole. On the inferred dipole, the great circles that intersect the dipole's poles all have the same, maximal $V_i$, while the equatorial and near-equatorial great circles have null $V_i$ and very small $V_i$, respectively. The highly sectoral $(\ell = 2)$ and $(\ell = 3)$ components aligned with the inferred dipole create variance along the equatorial and near-equatorial great circles, reducing variations of great-circle variance.\label{fig:equirectangular}}
\end{figure*}

\subsection{Same-spectrum realizations}

Let $\mathcal{M}$ be a CMB temperature map with angular power spectrum $\alpha_\ell$. We define a \emph{same-spectrum realization} of $\mathcal{M}$ as a CMB temperature map with angular power spectrum $\widetilde{\alpha}_\ell$ identical to the spectrum $\alpha_\ell$, and harmonic coefficients $\widetilde{\mathcal{A}}_{\ell m}$ generated in the following manner. For each $\ell$, one number is chosen from a normal distribution with zero mean and variance $\alpha_\ell$, and $2\ell$ numbers are chosen from a normal distribution with zero mean and variance $\alpha_\ell / 2$. Then, these numbers are respectively assigned to one real coefficient $\mathcal{A}_{\ell, 0}$ and $\ell$ complex coefficients $\mathcal{A}_{\ell m}$ for $m=1,2,\dots,\ell$, composed of $\ell$ real and $\ell$ imaginary components. By Eq. \eqref{eq:constr}, the coefficients $\mathcal{A}_{\ell m}$ for $m=-1,-2,\dots,-\ell$ are also determined. Finally, we define the $\widetilde{\mathcal{A}}_{\ell m}$ as
\begin{equation}
\widetilde{\mathcal{A}}_{\ell m} \equiv \mathcal{A}_{\ell m} \sqrt{\frac{\alpha_\ell(2\ell + 1)}{\sum_{m' = -\ell}^{m' = +\ell} \abs{\mathcal{A}_{\ell m'}}^2}},
\end{equation}
and these satisfy Eq. \eqref{eq:constr} by construction. Therefore, we have
\begin{align}
\widetilde{\alpha}_\ell &= \frac{1}{2\ell + 1}\sum_{m=-\ell}^{m=+\ell} \abs{\widetilde{\mathcal{A}}_{\ell m}}^2\\
&= \frac{1}{2\ell + 1} \cdot \frac{\alpha_\ell(2\ell + 1)}{\sum_{m' = -\ell}^{m' = +\ell} \abs{\mathcal{A}_{\ell m'}}^2} \cdot \sum_{m=-\ell}^{m=+\ell} \abs{\mathcal{A}_{\ell m}}^2\\
&= \alpha_\ell,
\end{align}
so both $\mathcal{M}$ and its same-spectrum realization have the same angular power spectrum (equivalently, two-point angular correlation function), while, in general, looking very different in temperature space.

\section{Method} \label{sec:method}

\subsection{Data}

Our analysis is based on foreground-corrected maps of the CMB temperature from the the third public release database of the \textsl{Planck} satellite. For this paper, we used the python wrapper for the Hierarchical Equal Area isoLatitude Pixelization (\textsl{HEALPix}) scheme \citep{Gorski_2005} on maps at a resolution defined by $N_{\text{side}} = 256$. We preprocessed the \textsl{Planck} maps by downgrading them to this resolution and removing their respective monopole and dipole spherical harmonic moments. Furthermore, we consider all \textsl{Planck} maps in parallel by conducting all measurements and operations on each \textsl{Planck} map independently. In each figure, we adopt the spread of all curves as an estimate of systematic error.

\subsection{Harmonic analysis}

To perform the spherical-harmonic decomposition of $\delta T$ for each $\mathcal{M}$, we used the \textsl{HEALPix} function \texttt{map2alm} to calculate the $a_{\ell m}$ components up to a cutoff of $\ell_{max} = 30$, which we used to compute the $C_\ell$ up to $\ell_{max}$ by Eq. \eqref{eq:cl}. Then, we determined $C(\Theta)$ by summing Eq. \eqref{eq:corr} up to the sharp cutoff $\ell_{max}$. In order to find $\hat{N}_\ell$, we used a differential evolution algorithm to search the space of all possible rotated coordinate systems.

\subsection{The Variance of Great-Circle Variances}

Let $X$ denote a sample of $n_s$ values taken with replacement from a distribution $Y$. The sample variance $s^2$ of $X$ is given by
\begin{equation} \label{eq:sv}
s^2(X) = \frac{1}{n_s - 1}\sum_{j=1}^{n_s} \left[X_j - \overline{X}\, \right]^2,
\end{equation}
where $\overline{X}$ is the sample mean,
\begin{equation}
\overline{X} = \frac{1}{n_s}\sum_{k=1}^{n_s} X_k.
\end{equation}
Since $s^2(X)$ is an unbiased estimator of $\sigma^2(Y)$, we have
\begin{equation} \label{eq:ev}
\ev{s^2(X)} = \sigma^2(Y),
\end{equation}
where $\ev{s^2(X)}$ is the expectation value of $s^2$ over all possible $X$ of same sample size $n_s$.

To calculate $V_i$, we first take a sample $(n_p = 2{,}000)$ of evenly spaced points $\vec{x}_j \in \mathbb{R}^3$ on $c_i$. Then, we use the \textsl{HEALPix} function \texttt{vec2pix} to map each $\vec{x}_j$ onto its corresponding pixel and subsequently, the value $\delta T(\vec{x}_j)$ of the CMB temperature deviation at that pixel. Next, we define the sample $A$ of $n_p$ values $A_j$ from the distribution $\delta T_i$ as
\begin{equation}
A_j = \delta T(\vec{x}_j),
\end{equation}
so $s^2(A)$ is an unbiased estimator of $\sigma^2(\delta T_i)$. Then, since $V_i = \sigma^2(\delta T_i)$, we use Eq. \eqref{eq:ev} to measure $V_i$ as $\ev{s^2(A)}$, which we compute as $s^2(A)$ from a single sample $A$ using Eq. \eqref{eq:sv}.

As a check on pixelation error, we calculated $V_i$ in the previously described manner for a set of $n_c = 20{,}000$ uniformly distributed random $c_i$ on one map at two different resolutions defined by $N_{\text{side}} = 256$ and $N_{\text{side}} = 128$, respectively. In all cases, the difference in the values of $V_i$ calculated at the different resolutions was less than $0.5\%$. Furthermore, to ensure sufficiently high sampling frequency of points from each great circle, we calculated $V_i$ in the previously described manner for a set of $n_c$ uniformly distributed random $c_i$ with samples of size $n_p$ and samples of size $n_p / 2$. For all great circles, the difference in the values of $V_i$ calculated with the different sample sizes was less than $0.4\%$. As a check on error due to beating between the pixelation pattern and the sampling frequency of points from each great circle, we calculated $V_i$ in the previously described manner for a set of $n_c$ uniformly distributed random $c_i$ with samples of size $n_p$ and with three other slightly reduced sample sizes $(n_p - 1, n_p - 2,$ and $n_p - 3$). For all great circles, the difference between the values of $V_i$ calculated with the samples of size $n_p$ and those calculated with any of the slightly reduced sample sizes was less than $0.3\%$.

Let $B$ denote a sample of $n_c$ great-circle variances $V_i$ from $V$, calculated from a set of $n_c$ uniformly distributed random $c_i$, so $s^2(B)$ is an unbiased estimator of $\sigma^2_V$. Then, since $\sigma^2_V=\sigma^2(V)$, we use Eq. \eqref{eq:ev} to measure $\sigma^2_V$ as $\ev{s^2(B)}$. For the inferred and dipole-free version of each $\ell \leqslant 3$ \textsl{Planck} map, we compute $\ev{s^2(B)}$ as the mean value of $s^2(B)$ for $100$ different samples $B$, where $s^2(B)$ is calculated using Eq. \eqref{eq:sv}; however, for $\ell \leqslant 3$ \textsl{Planck} maps that have had a standard model dipole $\mathcal{D}_\mathcal{S}$ added and for all other simulated realizations of the CMB sky, we measure $\ev{s^2(B)}$ as $s^2(B)$ from a single sample $B$, where $s^2(B)$ is calculated using Eq. \eqref{eq:sv}.

% As a check on the error from the finite number of great circles sampled, the spreads in the values of $s^2$ for the inferred and dipole-free version of each $\ell \leqslant 3$ \textsl{Planck} map are visible in the inset of Fig. \ref{fig:standard}.

\subsection{The inferred dipole}

To find $\mathcal{D}_\mathcal{I}(\mathcal{M})$, we use a differential evolution algorithm to search the space of all dipoles whose amplitudes are less than $3{,}000$ $\mu \text{K}^2$ for the $\mathcal{D}$ that minimizes $\sigma^2_V(\mathcal{M}_3' + \mathcal{D})$. We determined that, in practice, this space contains all possible values of $C_\mathcal{I}$ for the maps examined in this paper.

When calculating $\mathcal{D}_\mathcal{I}$, there is some error in each of the measurements of $\sigma^2_V$ made during differential evolution, which leads to some error in $\mathcal{D}_\mathcal{I}$. As a check on this error, $\mathcal{D}_\mathcal{I}$ was measured for each \textsl{Planck} map with $100$ different sets of $20{,}000$ uniformly distributed random great circles. In the inset of Fig. \ref{fig:amplitude}, we show the resulting spread of $C_\mathcal{I}$ values. Moreover, in all cases, the resulting spread in the value of $\hat{\Omega}_\mathcal{I}$ for each \textsl{Planck} map is less than $\pm 1.1^\circ$ of angular distance from the mean direction of $\hat{\Omega}_\mathcal{I}$ for that map. To create the inferred version of each \textsl{Planck} map, we use the mean values of $C_\mathcal{I}$ and $\hat{\Omega}_\mathcal{I}$ from this ensemble of calculations. For each simulated realization of the CMB sky, we perform single calculations of $\mathcal{D}_\mathcal{I}$ from one set of $20{,}000$ uniformly distributed random great circles to keep computational expense manageable.

In Fig \ref{fig:corrfunc}, the mean value of $C_\mathcal{I}$ for each \textsl{Planck} map is used to calculate the angular correlation function for each \textsl{Planck} map with its inferred dipole. The error in the calculation of $C_\mathcal{I}$ described in the previous paragraph propagates into the calculation of the angular correlation function for each \textsl{Planck} map with its inferred dipole; however, its effect on the correlation function is small: the angular correlation function calculated with any individual $C_\mathcal{I}$ differs from that calculated with the mean value of $C_\mathcal{I}$ by less than $3.5 \mu K^2$ at all points.

\subsection{Standard Model realizations}

We used the \emph{Code for Anisotropies in the Microwave Background} \citep{2011ascl.soft02026L} to calculate $C_\ell^\text{\,SM}$ with the following six cosmological parameters from the \textsl{Planck} collaboration \citep{2020A&A...641A...6P}: dark matter density $\Omega_c h^2 = 0.120$; baryon density $\Omega_b h^2 = 0.0224$; Hubble constant $H_0 = 67.4$; reionization optical depth $\tau = 0.054$; neutrino mass $m_\nu = 0.06$ eV; and spatial curvature $\Omega_k = 0.001$. For each standard model realization, we calculated the angular power spectrum up to a cutoff of $\ell_{max} = 30$ by Eq. \eqref{eq:cl}. Then, we determined $C(\Theta)$ by summing Eq. \eqref{eq:corr} up to the sharp cutoff $\ell_{max}$.

\section{Results}

\subsection{Two-point angular correlation function including the inferred dipole}

Fig. \ref{fig:corrfunc} shows how the addition of the dipole inferred from the posited great-circle variance symmetry causes $C(\Theta)$ to approximately vanish on the angular domain $\Theta = [90^\circ, 135^\circ]$, another symmetry predicted in some holographic quantum inflation scenarios. This agreement supports the consistency of the holographic hypothesis.

% The inferred dipole uncertainty in Fig. \ref{fig:corrfunc} is calculated in the following manner using the Commander \textsl{Planck} map; however, we have verified that this process produces virtually the same curve, regardless of the choice of particular \textsl{Planck} map. First, we measure $\mathcal{D}_\mathcal{I}$ $100$ times using $100$ different sets of $20{,}000$ uniformly distributed random great circles. This produces an ensemble of $100$ different values of $C_\mathcal{I}$, which allows us to calculate a mean value of $C_\mathcal{I}$. Additionally, there is a $C_\mathcal{I}$ that has the greatest absolute deviation from this mean value of $C_\mathcal{I}$. The inferred dipole uncertainty is calculated as the absolute value of the two-point angular correlation function calculated from the mean value of $C_\mathcal{I}$ minus the two-point angular correlation function calculated from the value of $C_\mathcal{I}$ that has the greatest absolute deviation from the mean value of $C_\mathcal{I}$.

In Fig. \ref{fig:amplitude}, we demonstrate that the amplitude $C_\mathcal{I}$ of the inferred dipole for same-spectrum realizations of our $\ell \leqslant 3$ sky is almost always lower than that measured for our real sky, and therefore, does not cause the vanishing of the two-point angular correlation function at large angles. For three of the maps, {\it all same-spectrum realizations that had higher or comparable inferred dipole amplitude to our sky had very nearly the same detailed characteristics as our sky: a highly sectoral octupole moment, and high alignment between the quadrupole and octupole}. (The exception is the outlier SEVEM, which had several realizations with less alignment between the quadrupole and octupole). Thus, our sky is extremely uncommon in that it is one of the few skies in which the symmetries predicted by the holographic model are self-consistent.

\subsection{Anomalous alignments and shapes in the natural coordinate system}

Fig. \ref{fig:mollweide} shows the low $\ell$ harmonic components of the inferred version of Commander in Galactic coordinates. Additionally, it demonstrates the close agreement between all four \textsl{Planck} maps on the inferred dipole axis $\hat{\Omega}_\mathcal{I}$, and the proximity of the octupole axis of maximum sectorality $\hat{N}_3$ to $\hat{\Omega}_\mathcal{I}$ for each \textsl{Planck} map.

In Fig. \ref{fig:equirectangular}, the equirectangular projections of the low $\ell$ harmonic components of the inferred version of Commander in its natural coordinate system illustrate the correlations between the octupole, quadrupole, and inferred dipole of our CMB sky. On the inferred dipole, the great circles that intersect the dipole's poles all have the same, maximal variance, while the equatorial and near-equatorial great circles have no variance and very little variance, respectively. On the quadrupole moment and the highly sectoral octupole moment, the great circles that intersect the poles all have relatively little variance, while the near-equatorial great circles have near-maximal variance. As a result, the quadrupole and octupole must be aligned with the inferred dipole and the octupole must be highly sectoral in order to create variance along the near-equatorial great circles and bring the $V_i$ closer to being constant, thus lowering $\sigma^2_V$. Equivalently, $\hat{\Omega}_\mathcal{I}$ must be close to the respective axes of maximum sectorality for the quadrupole and the octupole, and the octupole must have a high maximum value of $\mathcal{S}_3$.

\section{Conclusions}

The analysis presented here shows that several known anomalies of large-angle CMB anisotropy may be related to a universal symmetry of the primordial pattern of curvature perturbations, a causal universality of great circle variance on inflationary horizons. This simple geometrical symmetry connects several known properties in the CMB sky that are disconnected statistical anomalies in the standard model. We have shown that very few random combinations of independent $a_{\ell m}$ components drawn from the standard expected distribution simultaneously agree with the posited symmetry and the near-vanishing of angular correlation at angular separation $\Theta\geqslant 90^\circ$ better than the combination present in our sky.

Our results lend support to a new interpretation of long-studied large-angle CMB anomalies \cite{hogan2021angular}: several of them may not be independent random statistical flukes, but may be related signatures of new physics not currently included in the standard quantum system used to describe inflation. Even assuming a realization that agrees with the observed angular power spectrum, the results presented here show that the actual sky has an anomalous pattern, consistent with alignments and shapes of the lowest-order multipoles expected in a causally-coherent scenario with great-circle symmetry, that occurs in fewer than one percent of standard-scenario realizations. As indicated by the differences between the different maps, and particularly by the outlier SEVEM, more precise and powerful tests are currently limited by noise introduced by Galactic foreground models. One promising path forward to improve these would be better multi-frequency maps, preferably over the entire sky. New all sky experiments, in particular PIXIE \citep{kogut2011,Kogut2020} and LiteBIRD \citep{Litebird}, will provide all-sky CMB maps with well controlled large angular scale systematics and very broad spectral coverage to markedly reduce the uncertainties in the low-$\ell$ CMB structure.

\begin{acknowledgments}
This work was supported by the Fermi National Accelerator Laboratory (Fermilab), a U.S. Department of Energy, Office of Science, HEP User Facility, managed by Fermi Research Alliance, LLC (FRA), acting under Contract No. DE-AC02-07CH11359. We acknowledge the use of HEALPix/healpy and the NASA/IPAC Infrared Science Archive, which is operated by the Jet Propulsion Laboratory, California Institute of Technology, under contract with the National Aeronautics and Space Administration.
\end{acknowledgments}

\bibliography{universal}

\begin{thebibliography}{26}%
\makeatletter
\providecommand \@ifxundefined [1]{%
 \@ifx{#1\undefined}
}%
\providecommand \@ifnum [1]{%
 \ifnum #1\expandafter \@firstoftwo
 \else \expandafter \@secondoftwo
 \fi
}%
\providecommand \@ifx [1]{%
 \ifx #1\expandafter \@firstoftwo
 \else \expandafter \@secondoftwo
 \fi
}%
\providecommand \natexlab [1]{#1}%
\providecommand \enquote  [1]{``#1''}%
\providecommand \bibnamefont  [1]{#1}%
\providecommand \bibfnamefont [1]{#1}%
\providecommand \citenamefont [1]{#1}%
\providecommand \href@noop [0]{\@secondoftwo}%
\providecommand \href [0]{\begingroup \@sanitize@url \@href}%
\providecommand \@href[1]{\@@startlink{#1}\@@href}%
\providecommand \@@href[1]{\endgroup#1\@@endlink}%
\providecommand \@sanitize@url [0]{\catcode `\\12\catcode `\$12\catcode
  `\&12\catcode `\#12\catcode `\^12\catcode `\_12\catcode `\%12\relax}%
\providecommand \@@startlink[1]{}%
\providecommand \@@endlink[0]{}%
\providecommand \url  [0]{\begingroup\@sanitize@url \@url }%
\providecommand \@url [1]{\endgroup\@href {#1}{\urlprefix }}%
\providecommand \urlprefix  [0]{URL }%
\providecommand \Eprint [0]{\href }%
\providecommand \doibase [0]{http://dx.doi.org/}%
\providecommand \selectlanguage [0]{\@gobble}%
\providecommand \bibinfo  [0]{\@secondoftwo}%
\providecommand \bibfield  [0]{\@secondoftwo}%
\providecommand \translation [1]{[#1]}%
\providecommand \BibitemOpen [0]{}%
\providecommand \bibitemStop [0]{}%
\providecommand \bibitemNoStop [0]{.\EOS\space}%
\providecommand \EOS [0]{\spacefactor3000\relax}%
\providecommand \BibitemShut  [1]{\csname bibitem#1\endcsname}%
\let\auto@bib@innerbib\@empty
%</preamble>
\bibitem [{\citenamefont {{Planck
  Collaboration}}(2020{\natexlab{a}})}]{Akrami:2018vks}%
  \BibitemOpen
  \bibfield  {author} {\bibinfo {author} {\bibnamefont {{Planck
  Collaboration}}},\ }\bibfield  {title} {\enquote {\bibinfo {title} {{Planck
  2018 results. I. Overview and the cosmological legacy of Planck}},}\ }\href
  {\doibase 10.1051/0004-6361/201833880} {\bibfield  {journal} {\bibinfo
  {journal} {Astron. Astrophys.}\ }\textbf {\bibinfo {volume} {641}},\ \bibinfo
  {eid} {A1} (\bibinfo {year} {2020}{\natexlab{a}})},\ \Eprint
  {http://arxiv.org/abs/1807.06205} {arXiv:1807.06205 [astro-ph.CO]}
  \BibitemShut {NoStop}%
%%CITATION = ARXIV:1807.06205;%%
\bibitem [{\citenamefont {de~Oliveira-Costa}\ \emph {et~al.}(2004)\citenamefont
  {de~Oliveira-Costa}, \citenamefont {Tegmark}, \citenamefont {Zaldarriaga},\
  and\ \citenamefont {Hamilton}}]{de_oliveira-costa}%
  \BibitemOpen
  \bibfield  {author} {\bibinfo {author} {\bibfnamefont {Angelica}\
  \bibnamefont {de~Oliveira-Costa}}, \bibinfo {author} {\bibfnamefont {Max}\
  \bibnamefont {Tegmark}}, \bibinfo {author} {\bibfnamefont {Matias}\
  \bibnamefont {Zaldarriaga}}, \ and\ \bibinfo {author} {\bibfnamefont
  {Andrew}\ \bibnamefont {Hamilton}},\ }\bibfield  {title} {\enquote {\bibinfo
  {title} {The significance of the largest scale cmb fluctuations in wmap},}\
  }\href {\doibase 10.1103/PhysRevD.69.063516} {\bibfield  {journal} {\bibinfo
  {journal} {Physical Review D}\ }\textbf {\bibinfo {volume} {69}} (\bibinfo
  {year} {2004}),\ 10.1103/PhysRevD.69.063516}\BibitemShut {NoStop}%
\bibitem [{\citenamefont {Bennett}\ \emph {et~al.}(2011)\citenamefont
  {Bennett}, \citenamefont {Hill}, \citenamefont {Hinshaw}, \citenamefont
  {Larson}, \citenamefont {Smith}, \citenamefont {Dunkley}, \citenamefont
  {Gold}, \citenamefont {Halpern}, \citenamefont {Jarosik}, \citenamefont
  {Kogut}, \citenamefont {Komatsu}, \citenamefont {Limon}, \citenamefont
  {Meyer}, \citenamefont {Nolta}, \citenamefont {Odegard}, \citenamefont
  {Page}, \citenamefont {Spergel}, \citenamefont {Tucker}, \citenamefont
  {Weiland}, \citenamefont {Wollack},\ and\ \citenamefont
  {Wright}}]{WMAPanomalies}%
  \BibitemOpen
  \bibfield  {author} {\bibinfo {author} {\bibfnamefont {C.~L.}\ \bibnamefont
  {Bennett}}, \bibinfo {author} {\bibfnamefont {R.~S.}\ \bibnamefont {Hill}},
  \bibinfo {author} {\bibfnamefont {G.}~\bibnamefont {Hinshaw}}, \bibinfo
  {author} {\bibfnamefont {D.}~\bibnamefont {Larson}}, \bibinfo {author}
  {\bibfnamefont {K.~M.}\ \bibnamefont {Smith}}, \bibinfo {author}
  {\bibfnamefont {J.}~\bibnamefont {Dunkley}}, \bibinfo {author} {\bibfnamefont
  {B.}~\bibnamefont {Gold}}, \bibinfo {author} {\bibfnamefont {M.}~\bibnamefont
  {Halpern}}, \bibinfo {author} {\bibfnamefont {N.}~\bibnamefont {Jarosik}},
  \bibinfo {author} {\bibfnamefont {A.}~\bibnamefont {Kogut}}, \bibinfo
  {author} {\bibfnamefont {E.}~\bibnamefont {Komatsu}}, \bibinfo {author}
  {\bibfnamefont {M.}~\bibnamefont {Limon}}, \bibinfo {author} {\bibfnamefont
  {S.~S.}\ \bibnamefont {Meyer}}, \bibinfo {author} {\bibfnamefont {M.~R.}\
  \bibnamefont {Nolta}}, \bibinfo {author} {\bibfnamefont {N.}~\bibnamefont
  {Odegard}}, \bibinfo {author} {\bibfnamefont {L.}~\bibnamefont {Page}},
  \bibinfo {author} {\bibfnamefont {D.~N.}\ \bibnamefont {Spergel}}, \bibinfo
  {author} {\bibfnamefont {G.{\^A}~S.}\ \bibnamefont {Tucker}}, \bibinfo
  {author} {\bibfnamefont {J.~L.}\ \bibnamefont {Weiland}}, \bibinfo {author}
  {\bibfnamefont {E.}~\bibnamefont {Wollack}}, \ and\ \bibinfo {author}
  {\bibfnamefont {E.~L.}\ \bibnamefont {Wright}},\ }\bibfield  {title}
  {\enquote {\bibinfo {title} {Seven-year wilkinson microwave anisotropy probe
  (wmap) observations: Are there cosmic microwave background anomalies?}}\
  }\href {http://stacks.iop.org/0067-0049/192/i=2/a=17} {\bibfield  {journal}
  {\bibinfo  {journal} {The Astrophysical Journal Supplement Series}\ }\textbf
  {\bibinfo {volume} {192}},\ \bibinfo {pages} {17} (\bibinfo {year}
  {2011})}\BibitemShut {NoStop}%
\bibitem [{\citenamefont {{Planck Collaboration}}(2016)}]{2016A&A...594A..16P}%
  \BibitemOpen
  \bibfield  {author} {\bibinfo {author} {\bibnamefont {{Planck
  Collaboration}}},\ }\bibfield  {title} {\enquote {\bibinfo {title} {{Planck
  2015 results. XVI. Isotropy and statistics of the CMB}},}\ }\href {\doibase
  10.1051/0004-6361/201526681} {\bibfield  {journal} {\bibinfo  {journal}
  {Asro}\ }\textbf {\bibinfo {volume} {594}},\ \bibinfo {eid} {A16} (\bibinfo
  {year} {2016})},\ \Eprint {http://arxiv.org/abs/1506.07135} {arXiv:1506.07135
  [astro-ph.CO]} \BibitemShut {NoStop}%
\bibitem [{\citenamefont {{Schwarz}}\ \emph {et~al.}(2016)\citenamefont
  {{Schwarz}}, \citenamefont {{Copi}}, \citenamefont {{Huterer}},\ and\
  \citenamefont {{Starkman}}}]{2016CQGra..33r4001S}%
  \BibitemOpen
  \bibfield  {author} {\bibinfo {author} {\bibfnamefont {Dominik~J.}\
  \bibnamefont {{Schwarz}}}, \bibinfo {author} {\bibfnamefont {Craig~J.}\
  \bibnamefont {{Copi}}}, \bibinfo {author} {\bibfnamefont {Dragan}\
  \bibnamefont {{Huterer}}}, \ and\ \bibinfo {author} {\bibfnamefont
  {Glenn~D.}\ \bibnamefont {{Starkman}}},\ }\bibfield  {title} {\enquote
  {\bibinfo {title} {{CMB anomalies after Planck}},}\ }\href {\doibase
  10.1088/0264-9381/33/18/184001} {\bibfield  {journal} {\bibinfo  {journal}
  {Classical and Quantum Gravity}\ }\textbf {\bibinfo {volume} {33}},\ \bibinfo
  {eid} {184001} (\bibinfo {year} {2016})},\ \Eprint
  {http://arxiv.org/abs/1510.07929} {arXiv:1510.07929 [astro-ph.CO]}
  \BibitemShut {NoStop}%
\bibitem [{\citenamefont {{Planck
  Collaboration}}(2020{\natexlab{b}})}]{Planck-2018-Isotropy}%
  \BibitemOpen
  \bibfield  {author} {\bibinfo {author} {\bibnamefont {{Planck
  Collaboration}}},\ }\bibfield  {title} {\enquote {\bibinfo {title} {Planck
  2018 results: Vii. isotropy and statistics of the cmb},}\ }\href {\doibase
  https://doi.org/10.1051/0004-6361/201935201} {\bibfield  {journal} {\bibinfo
  {journal} {Astron. Astrophys.}\ }\textbf {\bibinfo {volume} {641}},\ \bibinfo
  {eid} {A7} (\bibinfo {year} {2020}{\natexlab{b}})}\BibitemShut {NoStop}%
\bibitem [{\citenamefont {{Wright}}\ \emph {et~al.}(1992)\citenamefont
  {{Wright}}, \citenamefont {{Meyer}}, \citenamefont {{Bennett}}, \citenamefont
  {{Boggess}}, \citenamefont {{Cheng}}, \citenamefont {{Hauser}}, \citenamefont
  {{Kogut}}, \citenamefont {{Lineweaver}}, \citenamefont {{Mather}},
  \citenamefont {{Smoot}}, \citenamefont {{Weiss}}, \citenamefont {{Gulkis}},
  \citenamefont {{Hinshaw}}, \citenamefont {{Janssen}}, \citenamefont
  {{Kelsall}}, \citenamefont {{Lubin}}, \citenamefont {{Moseley}},
  \citenamefont {{Murdock}}, \citenamefont {{Shafer}}, \citenamefont
  {{Silverberg}},\ and\ \citenamefont {{Wilkinson}}}]{1992ApJ...396L..13W}%
  \BibitemOpen
  \bibfield  {author} {\bibinfo {author} {\bibfnamefont {E.~L.}\ \bibnamefont
  {{Wright}}}, \bibinfo {author} {\bibfnamefont {S.~S.}\ \bibnamefont
  {{Meyer}}}, \bibinfo {author} {\bibfnamefont {C.~L.}\ \bibnamefont
  {{Bennett}}}, \bibinfo {author} {\bibfnamefont {N.~W.}\ \bibnamefont
  {{Boggess}}}, \bibinfo {author} {\bibfnamefont {E.~S.}\ \bibnamefont
  {{Cheng}}}, \bibinfo {author} {\bibfnamefont {M.~G.}\ \bibnamefont
  {{Hauser}}}, \bibinfo {author} {\bibfnamefont {A.}~\bibnamefont {{Kogut}}},
  \bibinfo {author} {\bibfnamefont {C.}~\bibnamefont {{Lineweaver}}}, \bibinfo
  {author} {\bibfnamefont {J.~C.}\ \bibnamefont {{Mather}}}, \bibinfo {author}
  {\bibfnamefont {G.~F.}\ \bibnamefont {{Smoot}}}, \bibinfo {author}
  {\bibfnamefont {R.}~\bibnamefont {{Weiss}}}, \bibinfo {author} {\bibfnamefont
  {S.}~\bibnamefont {{Gulkis}}}, \bibinfo {author} {\bibfnamefont
  {G.}~\bibnamefont {{Hinshaw}}}, \bibinfo {author} {\bibfnamefont
  {M.}~\bibnamefont {{Janssen}}}, \bibinfo {author} {\bibfnamefont
  {T.}~\bibnamefont {{Kelsall}}}, \bibinfo {author} {\bibfnamefont {P.~M.}\
  \bibnamefont {{Lubin}}}, \bibinfo {author} {\bibfnamefont {Jr.}\ \bibnamefont
  {{Moseley}}, \bibfnamefont {S.~H.}}, \bibinfo {author} {\bibfnamefont
  {T.~L.}\ \bibnamefont {{Murdock}}}, \bibinfo {author} {\bibfnamefont {R.~A.}\
  \bibnamefont {{Shafer}}}, \bibinfo {author} {\bibfnamefont {R.~F.}\
  \bibnamefont {{Silverberg}}}, \ and\ \bibinfo {author} {\bibfnamefont
  {D.~T.}\ \bibnamefont {{Wilkinson}}},\ }\bibfield  {title} {\enquote
  {\bibinfo {title} {{Interpretation of the Cosmic Microwave Background
  Radiation Anisotropy Detected by the COBE Differential Microwave
  Radiometer}},}\ }\href {\doibase 10.1086/186506} {\bibfield  {journal}
  {\bibinfo  {journal} {\apj}\ }\textbf {\bibinfo {volume} {396}},\ \bibinfo
  {pages} {L13} (\bibinfo {year} {1992})}\BibitemShut {NoStop}%
\bibitem [{\citenamefont {{Bennett}}\ \emph {et~al.}(1994)\citenamefont
  {{Bennett}}, \citenamefont {{Kogut}}, \citenamefont {{Hinshaw}},
  \citenamefont {{Banday}}, \citenamefont {{Wright}}, \citenamefont {{Gorski}},
  \citenamefont {{Wilkinson}}, \citenamefont {{Weiss}}, \citenamefont
  {{Smoot}}, \citenamefont {{Meyer}}, \citenamefont {{Mather}}, \citenamefont
  {{Lubin}}, \citenamefont {{Loewenstein}}, \citenamefont {{Lineweaver}},
  \citenamefont {{Keegstra}}, \citenamefont {{Kaita}}, \citenamefont
  {{Jackson}},\ and\ \citenamefont {{Cheng}}}]{1994ApJ...436..423B}%
  \BibitemOpen
  \bibfield  {author} {\bibinfo {author} {\bibfnamefont {C.~L.}\ \bibnamefont
  {{Bennett}}}, \bibinfo {author} {\bibfnamefont {A.}~\bibnamefont {{Kogut}}},
  \bibinfo {author} {\bibfnamefont {G.}~\bibnamefont {{Hinshaw}}}, \bibinfo
  {author} {\bibfnamefont {A.~J.}\ \bibnamefont {{Banday}}}, \bibinfo {author}
  {\bibfnamefont {E.~L.}\ \bibnamefont {{Wright}}}, \bibinfo {author}
  {\bibfnamefont {K.~M.}\ \bibnamefont {{Gorski}}}, \bibinfo {author}
  {\bibfnamefont {D.~T.}\ \bibnamefont {{Wilkinson}}}, \bibinfo {author}
  {\bibfnamefont {R.}~\bibnamefont {{Weiss}}}, \bibinfo {author} {\bibfnamefont
  {G.~F.}\ \bibnamefont {{Smoot}}}, \bibinfo {author} {\bibfnamefont {S.~S.}\
  \bibnamefont {{Meyer}}}, \bibinfo {author} {\bibfnamefont {J.~C.}\
  \bibnamefont {{Mather}}}, \bibinfo {author} {\bibfnamefont {P.}~\bibnamefont
  {{Lubin}}}, \bibinfo {author} {\bibfnamefont {K.}~\bibnamefont
  {{Loewenstein}}}, \bibinfo {author} {\bibfnamefont {C.}~\bibnamefont
  {{Lineweaver}}}, \bibinfo {author} {\bibfnamefont {P.}~\bibnamefont
  {{Keegstra}}}, \bibinfo {author} {\bibfnamefont {E.}~\bibnamefont {{Kaita}}},
  \bibinfo {author} {\bibfnamefont {P.~D.}\ \bibnamefont {{Jackson}}}, \ and\
  \bibinfo {author} {\bibfnamefont {E.~S.}\ \bibnamefont {{Cheng}}},\
  }\bibfield  {title} {\enquote {\bibinfo {title} {{Cosmic Temperature
  Fluctuations from Two Years of COBE Differential Microwave Radiometers
  Observations}},}\ }\href {\doibase 10.1086/174918} {\bibfield  {journal}
  {\bibinfo  {journal} {\apj}\ }\textbf {\bibinfo {volume} {436}},\ \bibinfo
  {pages} {423} (\bibinfo {year} {1994})},\ \Eprint
  {http://arxiv.org/abs/astro-ph/9401012} {arXiv:astro-ph/9401012 [astro-ph]}
  \BibitemShut {NoStop}%
\bibitem [{\citenamefont {Copi}\ \emph {et~al.}(2010)\citenamefont {Copi},
  \citenamefont {Huterer}, \citenamefont {Schwarz},\ and\ \citenamefont
  {Starkman}}]{Copi2010}%
  \BibitemOpen
  \bibfield  {author} {\bibinfo {author} {\bibfnamefont {Craig~J.}\
  \bibnamefont {Copi}}, \bibinfo {author} {\bibfnamefont {Dragan}\ \bibnamefont
  {Huterer}}, \bibinfo {author} {\bibfnamefont {Dominik~J.}\ \bibnamefont
  {Schwarz}}, \ and\ \bibinfo {author} {\bibfnamefont {Glenn~D.}\ \bibnamefont
  {Starkman}},\ }\bibfield  {title} {\enquote {\bibinfo {title} {Large-angle
  anomalies in the cmb},}\ }\href {\doibase 10.1155/2010/847541} {\bibfield
  {journal} {\bibinfo  {journal} {Advances in Astronomy}\ }\textbf {\bibinfo
  {volume} {2010}},\ \bibinfo {pages} {847541} (\bibinfo {year}
  {2010})}\BibitemShut {NoStop}%
\bibitem [{\citenamefont {Hagimoto}\ \emph {et~al.}(2020)\citenamefont
  {Hagimoto}, \citenamefont {Hogan}, \citenamefont {Lewin},\ and\ \citenamefont
  {Meyer}}]{Hagimoto_2020}%
  \BibitemOpen
  \bibfield  {author} {\bibinfo {author} {\bibfnamefont {Ray}\ \bibnamefont
  {Hagimoto}}, \bibinfo {author} {\bibfnamefont {Craig}\ \bibnamefont {Hogan}},
  \bibinfo {author} {\bibfnamefont {Collin}\ \bibnamefont {Lewin}}, \ and\
  \bibinfo {author} {\bibfnamefont {Stephan~S.}\ \bibnamefont {Meyer}},\
  }\bibfield  {title} {\enquote {\bibinfo {title} {Symmetries of {CMB}
  temperature correlation at large angular separations},}\ }\href {\doibase
  10.3847/2041-8213/ab62a0} {\bibfield  {journal} {\bibinfo  {journal} {The
  Astrophysical Journal}\ }\textbf {\bibinfo {volume} {888}},\ \bibinfo {pages}
  {L29} (\bibinfo {year} {2020})}\BibitemShut {NoStop}%
\bibitem [{\citenamefont {{Copi}}\ \emph {et~al.}(2015)\citenamefont {{Copi}},
  \citenamefont {{Huterer}}, \citenamefont {{Schwarz}},\ and\ \citenamefont
  {{Starkman}}}]{2015MNRAS.449.3458C}%
  \BibitemOpen
  \bibfield  {author} {\bibinfo {author} {\bibfnamefont {Craig~J.}\
  \bibnamefont {{Copi}}}, \bibinfo {author} {\bibfnamefont {Dragan}\
  \bibnamefont {{Huterer}}}, \bibinfo {author} {\bibfnamefont {Dominik~J.}\
  \bibnamefont {{Schwarz}}}, \ and\ \bibinfo {author} {\bibfnamefont
  {Glenn~D.}\ \bibnamefont {{Starkman}}},\ }\bibfield  {title} {\enquote
  {\bibinfo {title} {{Large-scale alignments from WMAP and Planck}},}\ }\href
  {\doibase 10.1093/mnras/stv501} {\bibfield  {journal} {\bibinfo  {journal}
  {MNRAS}\ }\textbf {\bibinfo {volume} {449}},\ \bibinfo {pages} {3458--3470}
  (\bibinfo {year} {2015})},\ \Eprint {http://arxiv.org/abs/1311.4562}
  {arXiv:1311.4562 [astro-ph.CO]} \BibitemShut {NoStop}%
\bibitem [{\citenamefont {Kim}\ and\ \citenamefont
  {Naselsky}(2010)}]{PhysRevD.82.063002}%
  \BibitemOpen
  \bibfield  {author} {\bibinfo {author} {\bibfnamefont {Jaiseung}\
  \bibnamefont {Kim}}\ and\ \bibinfo {author} {\bibfnamefont {Pavel}\
  \bibnamefont {Naselsky}},\ }\bibfield  {title} {\enquote {\bibinfo {title}
  {Anomalous parity asymmetry of wmap 7-year power spectrum data at low
  multipoles: Is it cosmological or systematics?}}\ }\href {\doibase
  10.1103/PhysRevD.82.063002} {\bibfield  {journal} {\bibinfo  {journal} {Phys.
  Rev. D}\ }\textbf {\bibinfo {volume} {82}},\ \bibinfo {pages} {063002}
  (\bibinfo {year} {2010})}\BibitemShut {NoStop}%
\bibitem [{\citenamefont {Hogan}(2019)}]{PhysRevD.99.063531}%
  \BibitemOpen
  \bibfield  {author} {\bibinfo {author} {\bibfnamefont {Craig}\ \bibnamefont
  {Hogan}},\ }\bibfield  {title} {\enquote {\bibinfo {title} {Nonlocal
  entanglement and directional correlations of primordial perturbations on the
  inflationary horizon},}\ }\href {\doibase 10.1103/PhysRevD.99.063531}
  {\bibfield  {journal} {\bibinfo  {journal} {Phys. Rev. D}\ }\textbf {\bibinfo
  {volume} {99}},\ \bibinfo {pages} {063531} (\bibinfo {year}
  {2019})}\BibitemShut {NoStop}%
\bibitem [{\citenamefont {Hogan}(2020)}]{Hogan_2020}%
  \BibitemOpen
  \bibfield  {author} {\bibinfo {author} {\bibfnamefont {Craig}\ \bibnamefont
  {Hogan}},\ }\bibfield  {title} {\enquote {\bibinfo {title} {Pattern of
  perturbations from a coherent quantum inflationary horizon},}\ }\href
  {\doibase 10.1088/1361-6382/ab7964} {\bibfield  {journal} {\bibinfo
  {journal} {Classical and Quantum Gravity}\ }\textbf {\bibinfo {volume}
  {37}},\ \bibinfo {pages} {095005} (\bibinfo {year} {2020})}\BibitemShut
  {NoStop}%
\bibitem [{\citenamefont {Hogan}\ and\ \citenamefont
  {Meyer}(2021)}]{hogan2021angular}%
  \BibitemOpen
  \bibfield  {author} {\bibinfo {author} {\bibfnamefont {Craig}\ \bibnamefont
  {Hogan}}\ and\ \bibinfo {author} {\bibfnamefont {Stephan~S.}\ \bibnamefont
  {Meyer}},\ }\bibfield  {title} {\enquote {\bibinfo {title} {Angular
  correlations of causally-coherent primordial quantum perturbations},}\
  }\href@noop {} {\  (\bibinfo {year} {2021})},\ \Eprint
  {http://arxiv.org/abs/2109.11092} {arXiv:2109.11092 [gr-qc]} \BibitemShut
  {NoStop}%
\bibitem [{\citenamefont {Banks}\ and\ \citenamefont
  {Fischler}(2018)}]{Banks:2018ypk}%
  \BibitemOpen
  \bibfield  {author} {\bibinfo {author} {\bibfnamefont {Tom}\ \bibnamefont
  {Banks}}\ and\ \bibinfo {author} {\bibfnamefont {W.}~\bibnamefont
  {Fischler}},\ }\bibfield  {title} {\enquote {\bibinfo {title} {{The
  holographic spacetime model of cosmology}},}\ }\href {\doibase
  10.1142/S0218271818460057} {\bibfield  {journal} {\bibinfo  {journal} {Int.
  J. Mod. Phys.}\ }\textbf {\bibinfo {volume} {D27}},\ \bibinfo {pages}
  {1846005} (\bibinfo {year} {2018})},\ \Eprint
  {http://arxiv.org/abs/1806.01749} {arXiv:1806.01749 [hep-th]} \BibitemShut
  {NoStop}%
%%CITATION = ARXIV:1806.01749;%%
\bibitem [{\citenamefont {Banks}(2020)}]{Banks:2020dus}%
  \BibitemOpen
  \bibfield  {author} {\bibinfo {author} {\bibfnamefont {Tom}\ \bibnamefont
  {Banks}},\ }\bibfield  {title} {\enquote {\bibinfo {title} {{Holographic
  Space-time and Quantum Information}},}\ }\href {\doibase
  10.3389/fphy.2020.00111} {\bibfield  {journal} {\bibinfo  {journal} {Front.
  in Phys.}\ }\textbf {\bibinfo {volume} {8}},\ \bibinfo {pages} {111}
  (\bibinfo {year} {2020})},\ \Eprint {http://arxiv.org/abs/2001.08205}
  {arXiv:2001.08205 [hep-th]} \BibitemShut {NoStop}%
\bibitem [{\citenamefont {Banks}\ and\ \citenamefont
  {Zurek}(2021)}]{Banks:2021jwj}%
  \BibitemOpen
  \bibfield  {author} {\bibinfo {author} {\bibfnamefont {Thomas}\ \bibnamefont
  {Banks}}\ and\ \bibinfo {author} {\bibfnamefont {Kathryn~M.}\ \bibnamefont
  {Zurek}},\ }\bibfield  {title} {\enquote {\bibinfo {title} {{Conformal
  Description of Near-Horizon Vacuum States}},}\ }\href@noop {} {\  (\bibinfo
  {year} {2021})},\ \Eprint {http://arxiv.org/abs/2108.04806} {arXiv:2108.04806
  [hep-th]} \BibitemShut {NoStop}%
\bibitem [{\citenamefont {Ferreira}\ and\ \citenamefont
  {Quartin}(2021)}]{Ferreira_2021}%
  \BibitemOpen
  \bibfield  {author} {\bibinfo {author} {\bibfnamefont {Pedro da~Silveira}\
  \bibnamefont {Ferreira}}\ and\ \bibinfo {author} {\bibfnamefont {Miguel}\
  \bibnamefont {Quartin}},\ }\bibfield  {title} {\enquote {\bibinfo {title}
  {First constraints on the intrinsic cmb dipole and our velocity with doppler
  and aberration},}\ }\href {\doibase 10.1103/PhysRevLett.127.101301}
  {\bibfield  {journal} {\bibinfo  {journal} {Phys. Rev. Lett.}\ }\textbf
  {\bibinfo {volume} {127}},\ \bibinfo {pages} {101301} (\bibinfo {year}
  {2021})}\BibitemShut {NoStop}%
\bibitem [{\citenamefont {Nadolny}\ \emph {et~al.}(2021)\citenamefont
  {Nadolny}, \citenamefont {Durrer}, \citenamefont {Kunz},\ and\ \citenamefont
  {Padmanabhan}}]{Nadolny_2021}%
  \BibitemOpen
  \bibfield  {author} {\bibinfo {author} {\bibfnamefont {Tobias}\ \bibnamefont
  {Nadolny}}, \bibinfo {author} {\bibfnamefont {Ruth}\ \bibnamefont {Durrer}},
  \bibinfo {author} {\bibfnamefont {Martin}\ \bibnamefont {Kunz}}, \ and\
  \bibinfo {author} {\bibfnamefont {Hamsa}\ \bibnamefont {Padmanabhan}},\
  }\bibfield  {title} {\enquote {\bibinfo {title} {A new way to test the
  cosmological principle: measuring our peculiar velocity and the large-scale
  anisotropy independently},}\ }\href {\doibase 10.1088/1475-7516/2021/11/009}
  {\bibfield  {journal} {\bibinfo  {journal} {Journal of Cosmology and
  Astroparticle Physics}\ }\textbf {\bibinfo {volume} {2021}},\ \bibinfo
  {pages} {009} (\bibinfo {year} {2021})}\BibitemShut {NoStop}%
\bibitem [{\citenamefont {Gorski}\ \emph {et~al.}(2005)\citenamefont {Gorski},
  \citenamefont {Hivon}, \citenamefont {Banday}, \citenamefont {Wandelt},
  \citenamefont {Hansen}, \citenamefont {Reinecke},\ and\ \citenamefont
  {Bartelmann}}]{Gorski_2005}%
  \BibitemOpen
  \bibfield  {author} {\bibinfo {author} {\bibfnamefont {K.~M.}\ \bibnamefont
  {Gorski}}, \bibinfo {author} {\bibfnamefont {E.}~\bibnamefont {Hivon}},
  \bibinfo {author} {\bibfnamefont {A.~J.}\ \bibnamefont {Banday}}, \bibinfo
  {author} {\bibfnamefont {B.~D.}\ \bibnamefont {Wandelt}}, \bibinfo {author}
  {\bibfnamefont {F.~K.}\ \bibnamefont {Hansen}}, \bibinfo {author}
  {\bibfnamefont {M.}~\bibnamefont {Reinecke}}, \ and\ \bibinfo {author}
  {\bibfnamefont {M.}~\bibnamefont {Bartelmann}},\ }\bibfield  {title}
  {\enquote {\bibinfo {title} {{HEALPix}: A framework for high-resolution
  discretization and fast analysis of data distributed on the sphere},}\ }\href
  {\doibase 10.1086/427976} {\bibfield  {journal} {\bibinfo  {journal} {The
  Astrophysical Journal}\ }\textbf {\bibinfo {volume} {622}},\ \bibinfo {pages}
  {759--771} (\bibinfo {year} {2005})}\BibitemShut {NoStop}%
\bibitem [{\citenamefont {{Lewis}}\ and\ \citenamefont
  {{Challinor}}(2011)}]{2011ascl.soft02026L}%
  \BibitemOpen
  \bibfield  {author} {\bibinfo {author} {\bibfnamefont {Antony}\ \bibnamefont
  {{Lewis}}}\ and\ \bibinfo {author} {\bibfnamefont {Anthony}\ \bibnamefont
  {{Challinor}}},\ }\href@noop {} {\enquote {\bibinfo {title} {{CAMB: Code for
  Anisotropies in the Microwave Background}},}\ } (\bibinfo {year} {2011}),\
  \Eprint {http://arxiv.org/abs/1102.026} {ascl:1102.026} \BibitemShut
  {NoStop}%
\bibitem [{\citenamefont {{Planck
  Collaboration}}(2020{\natexlab{c}})}]{2020A&A...641A...6P}%
  \BibitemOpen
  \bibfield  {author} {\bibinfo {author} {\bibnamefont {{Planck
  Collaboration}}},\ }\bibfield  {title} {\enquote {\bibinfo {title} {{Planck
  2018 results. VI. Cosmological parameters}},}\ }\href {\doibase
  10.1051/0004-6361/201833910} {\bibfield  {journal} {\bibinfo  {journal}
  {Astron. Astrophys.}\ }\textbf {\bibinfo {volume} {641}},\ \bibinfo {eid}
  {A6} (\bibinfo {year} {2020}{\natexlab{c}})},\ \Eprint
  {http://arxiv.org/abs/1807.06209} {arXiv:1807.06209 [astro-ph.CO]}
  \BibitemShut {NoStop}%
\bibitem [{\citenamefont {{Kogut}}\ \emph {et~al.}(2011)\citenamefont
  {{Kogut}}, \citenamefont {{Fixsen}}, \citenamefont {{Chuss}}, \citenamefont
  {{Dotson}}, \citenamefont {{Dwek}}, \citenamefont {{Halpern}}, \citenamefont
  {{Hinshaw}}, \citenamefont {{Meyer}}, \citenamefont {{Moseley}},
  \citenamefont {{Seiffert}}, \citenamefont {{Spergel}},\ and\ \citenamefont
  {{Wollack}}}]{kogut2011}%
  \BibitemOpen
  \bibfield  {author} {\bibinfo {author} {\bibfnamefont {A.}~\bibnamefont
  {{Kogut}}}, \bibinfo {author} {\bibfnamefont {D.~J.}\ \bibnamefont
  {{Fixsen}}}, \bibinfo {author} {\bibfnamefont {D.~T.}\ \bibnamefont
  {{Chuss}}}, \bibinfo {author} {\bibfnamefont {J.}~\bibnamefont {{Dotson}}},
  \bibinfo {author} {\bibfnamefont {E.}~\bibnamefont {{Dwek}}}, \bibinfo
  {author} {\bibfnamefont {M.}~\bibnamefont {{Halpern}}}, \bibinfo {author}
  {\bibfnamefont {G.~F.}\ \bibnamefont {{Hinshaw}}}, \bibinfo {author}
  {\bibfnamefont {S.~M.}\ \bibnamefont {{Meyer}}}, \bibinfo {author}
  {\bibfnamefont {S.~H.}\ \bibnamefont {{Moseley}}}, \bibinfo {author}
  {\bibfnamefont {M.~D.}\ \bibnamefont {{Seiffert}}}, \bibinfo {author}
  {\bibfnamefont {D.~N.}\ \bibnamefont {{Spergel}}}, \ and\ \bibinfo {author}
  {\bibfnamefont {E.~J.}\ \bibnamefont {{Wollack}}},\ }\bibfield  {title}
  {\enquote {\bibinfo {title} {{The Primordial Inflation Explorer (PIXIE): a
  nulling polarimeter for cosmic microwave background observations}},}\ }\href
  {\doibase 10.1088/1475-7516/2011/07/025} {\bibfield  {journal} {\bibinfo
  {journal} {JCAP}\ }\textbf {\bibinfo {volume} {2011}},\ \bibinfo {eid} {025}
  (\bibinfo {year} {2011})},\ \Eprint {http://arxiv.org/abs/1105.2044}
  {arXiv:1105.2044 [astro-ph.CO]} \BibitemShut {NoStop}%
\bibitem [{\citenamefont {{Kogut}}\ and\ \citenamefont
  {{Fixsen}}(2020)}]{Kogut2020}%
  \BibitemOpen
  \bibfield  {author} {\bibinfo {author} {\bibfnamefont {A.}~\bibnamefont
  {{Kogut}}}\ and\ \bibinfo {author} {\bibfnamefont {D.~J.}\ \bibnamefont
  {{Fixsen}}},\ }\bibfield  {title} {\enquote {\bibinfo {title} {{Calibration
  method and uncertainty for the primordial inflation explorer (PIXIE)}},}\
  }\href {\doibase 10.1088/1475-7516/2020/05/041} {\bibfield  {journal}
  {\bibinfo  {journal} {JCAP}\ }\textbf {\bibinfo {volume} {2020}},\ \bibinfo
  {eid} {041} (\bibinfo {year} {2020})},\ \Eprint
  {http://arxiv.org/abs/2002.00976} {arXiv:2002.00976 [astro-ph.IM]}
  \BibitemShut {NoStop}%
\bibitem [{\citenamefont {Sugai}(2020)}]{Litebird}%
  \BibitemOpen
  \bibfield  {author} {\bibinfo {author} {\bibfnamefont {H.}~\bibnamefont
  {Sugai}},\ }\bibfield  {title} {\enquote {\bibinfo {title} {Updated design of
  the cmb polarization experiment satellite litebird},}\ }\href@noop {}
  {\bibfield  {journal} {\bibinfo  {journal} {Journal of Low Temperature
  Physics}\ }\textbf {\bibinfo {volume} {199}},\ \bibinfo {pages} {1107--1117}
  (\bibinfo {year} {2020})}\BibitemShut {NoStop}%
\end{thebibliography}%

\end{document}